\documentclass[12pt,preprint]{aastex}

\newcommand{\om}{\ensuremath{\Omega_m}}

\newcommand{\ola}{\ensuremath{\Omega_{\Lambda}}}
\newcommand{\kms}{{\ensuremath{{\rm km~s}^{-1}}}}
\newcommand{\kmsmpc}{{\ensuremath{{\rm km~s}^{-1}~{\rm Mpc}^{-1}}}}

\newcommand{\cosmol}[3]{\ensuremath{\om = #1,\;\ola = #2,\;H_0 = #3~\kmsmpc}}
\newcommand{\lcdmparm}{\cosmol{0.27}{0.73}{71}}
\newcommand{\etal}{et~al.\/}

\newcommand{\hi}{\ion{H}{1}}
\newcommand{\msun}{\ensuremath{{\rm M}_\sun}}

\newcommand{\dmstd}[4]{\ensuremath{{#1}\degr\,{#2}\arcmin\,{#3}\farcs{#4}}}
\newcommand{\hmstd}[4]{\ensuremath{{#1}^{\rm h}\,{#2}^{\rm m}\,{#3}\fs{#4}}}

\newcommand{\dmsd}[4]{\ensuremath{{#1}\ {#2}\ {#3}.{#4}}}
\newcommand{\hmsd}[4]{\ensuremath{{#1}\ {#2}\ {#3}.{#4}}}

\newcommand{\ujy}{\ensuremath{\mu}Jy}

\newcommand{\ujypbm}{\ensuremath{\mu}{\rm Jy~beam}\ensuremath{^{-1}}}

\newcommand{\ie}{i.\,e.}
\newcommand{\eg}{e.\,g.}

\newcommand{\esc}{erg s\ensuremath{^{-1}} cm\ensuremath{^{-2}}}
\newcommand{\esec}{\ensuremath{{\rm erg~s}^{-1}}}
\newcommand{\esh}{\ensuremath{{\rm erg~s}^{-1}\,{\rm Hz}^{-1}}}

\newcommand{\ks}{\ensuremath{K_s}}

\newcommand{\chandra}{{\em Chandra}}

\newcommand{\chisq}{\ensuremath{\chi^2}}
\newcommand{\rchisq}{\ensuremath{\chi^2_\nu}}

\newcommand{\fullband}{\ensuremath{0.3 - 10}\,keV}

\shorttitle{X-ray Counterparts to Radio Transients}
\shortauthors{Croft \etal}

\begin{document}
\title{X-ray Observations of Radio Transients without Optical Hosts}
\author{Steve Croft\altaffilmark{1}, John A.\ Tomsick\altaffilmark{2}, and Geoffrey C.\ Bower\altaffilmark{1}}
\altaffiltext{1}{Astronomy Department, University of California, Berkeley, 601 Campbell Hall \#3411, Berkeley, CA 94720-3411, USA}
\altaffiltext{2}{Space Sciences Laboratory, 7 Gauss Way, University of California, Berkeley, CA 94720-7450, USA}
\tabletypesize{\scriptsize}

\begin{abstract}

We present a 50\,ks \chandra\ ACIS-I X-ray observation of the \citeauthor{bower} VLA archival field. The observations reach a limiting sensitivity of $\sim 10^{-4}$~counts~s$^{-1}$, corresponding to a flux of a few times $10^{-15}$\,\esc\ for the models we explore. The \chandra\ observations were undertaken to search for X-ray counterparts to the eight transient sources without optical counterparts, and the two transient sources with optical counterparts seen by \citeauthor{bower} Neither of the sources with optical counterparts was detected in X-rays. One of the eight optical non-detections is associated with a marginal ($2.4\sigma$) X-ray detection in our \chandra\ image. A second optically-undetected \citeauthor{bower} transient may be associated with a $z = 1.29$ X-ray detected quasar or its host galaxy, or alternatively is undetected in X-rays and is a chance association with the nearby X-ray source. The X-ray flux upper limits, and the one marginal detection, are consistent with the interpretation of \citeauthor{ofek:10} that the optically-undetected radio transients are flares from isolated old Galactic neutron stars. The marginal X-ray detection has a hardness ratio which implies a temperature too high for a simple one-temperature neutron star model, but plausible multi-component fits are not excluded, and in any case the marginal X-ray detection may be due to cosmic rays or particle background. The X-ray flux upper limits are also consistent with flare star progenitors at $\gtrsim 1$\,kpc (which would require the radio luminosity of the transient to be unusually high for such an object) or less extreme flares from brown dwarfs at distances of around 100\,pc.
\end{abstract}


\section{Introduction}

In recent years, authors such as \citet{bower}, hereafter B07; \citet{becker:10}; \citet{bower:11}; \citet{bannister:11}; \cite{bell:11}; and others, have made use of archival data (observations of calibration fields, or in some cases, entire telescope archives) to search for radio transients. Other studies have targeted particular fields with observations dedicated to the search for transients \citep[][and references therein]{ofek:11}.

Such studies are preparing the groundwork for a new generation of high-throughput radio telescopes: both dedicated new observatories --- for example, the Allen Telescope Array \citep{welch}, LOFAR \citep{lofar}, and ASKAP \citep{askap} --- and upgrades to older instruments --- for example, Apertif \citep{apertif} and EVLA \citep{evla}. A major goal for these new facilities is to survey large areas of sky at high sensitivity, both to build up deep, wide-field images, and to search for transient and variable sources from a range of progenitors \citep{lazio}. 

Currently, large areas of rate versus sensitivity parameter space remain to be explored \citep{pigss,paperii}, and the radio transient population is not well understood. In observations where radio transients are detected, sometimes counterparts are seen in existing surveys at other wavelengths. For example, transients may coincide with the nuclear regions of galaxies, in which case (depending on the galaxy redshift and hence the inferred radio luminosity of the transients) they are most likely due to active galactic nuclei (AGNs), or perhaps supernovae (SNe) or gamma ray bursts (GRBs). Or they may be offset from the peak of the optical emission, as is the case for two of the transients seen by B07, leading to their interpretation as SNe or GRBs.

For some radio transients, however, no counterparts are seen in archival images at other wavelengths, and no follow-up observations taken close in time to when the transients were seen are available. Follow-up observations will become more common as radio observatories move increasingly into a regime of issuing transient alerts for rapid follow-up by other facilities, but the field is still in its infancy compared to optical transient facilities which are now issuing transient alerts as a matter of routine \citep[\eg,][]{ptf}. However, the lack of galaxy counterparts in deep optical images can help exclude some models for these sources (such as AGNs, SNe and GRBs, unless these sources are at very high redshift or extremely obscured), and the lack of stellar counterparts can also constrain models with Galactic progenitors for these events. B07 consider radio SNe, GRBs, late-type stars, soft gamma-ray repeaters, X-ray binaries, pulsars, microlensing, and reflected solar flares as possible progenitors. They conclude that late-type stars are the most likely progenitors for transients with mJy radio fluxes and no optical counterparts. 

\citet{ofek:10}, hereafter O10, suggest old isolated Galactic neutron stars as a possible progenitor for the sources without optical identifications, and \citet{nakar:11} suggest that the sources with optical counterparts may be due to neutron star -- neutron star or neutron star -- black hole mergers. The latter are of particular interest because they are expected to also be producers of gravitational waves. X-ray observations 
can be a powerful discriminant among progenitor models, but information on the X-ray properties of the transients and transient hosts
is limited.  With this in mind, we obtained \chandra\ observations of the B07 field.

We assume an \lcdmparm\ cosmology \citep{wmap}. Magnitudes are given in the Vega system unless stated otherwise.

\section{Observations}

\subsection{Radio, optical, and Infra-Red Observations of the \citet{bower} field}

B07 analyzed archival data at 5 and 8.4\,GHz of a Very Large Array (VLA) calibration field centered at \hmstd{15}{02}{20}{53}, \dmstd{78}{16}{14}{905}. The data spanned $\sim 20$\,yr with 944 epochs, approximately one per week, each with an integration time of $\sim 20$\,min, all on the same field. The images had a full-width at half-maximum (FWHM) of 8\farcm6 and 5\farcm1 at 5 and 8.4\,GHz respectively, and typical root-mean-square (rms) fluctuations at the center of the images were $\sim 50$\,\ujy\ per epoch. 

B07 compared the images from epoch to epoch, and detected 7 single-epoch transient sources at 5\,GHz, and 1 at 8.4\,GHz. The transients are listed in Table~\ref{tab:radio}. B07 denoted these sources with identifiers beginning ``RT'' and the epoch in which they were observed; for convenience and clarity here we denote the 5\,GHz and 8\,GHz transients with identifiers beginning ``5T'' and ``8T'' respectively, although their formal identifiers are those beginning ``RT'' with an epoch of observation as cross-referenced in Table~\ref{tab:radio}.
For each of these transients, there was no simultaneous coverage at the other of the two observing frequencies, and no detection of the sources in longer-timescale average images.

\begin{deluxetable}{lllllllll}
\rotate
\tablewidth{0pt}
\tabletypesize{\scriptsize}
\tablecaption{\label{tab:radio} VLA sources and their \chandra\ counterparts}
\tablehead {
\colhead{VLA ID} & 
\colhead{VLA RA\tablenotemark{a}} &
\colhead{VLA Dec\tablenotemark{a}} &
\colhead{\chandra} &
\colhead{ACIS} &
\colhead{0.3 -- 10 keV rate\tablenotemark{d}} &
\colhead{Hardness} &
\colhead{Offset\tablenotemark{f}} &
\colhead{Optical / IR}\\
 & 
\colhead{(J2000)} &
\colhead{(J2000)} &
ID\tablenotemark{b} &
\colhead{counts\tablenotemark{c,d}} &
\colhead{(counts s$^{-1}$)} &
\colhead{constraint\tablenotemark{e}} &
\colhead{(arcsec)} &
\colhead{ID\tablenotemark{g}}
}
\startdata
5S1                              & \hmsd{15}{00}{11}{48} $\pm$ 0.17& \dmsd{78}{12}{42}{3} $\pm$ 0.2& X29       & $167 \pm 21$      & $3.38 \times 10^{-3}$     & $-0.62 \pm 0.11$ & $10.79 \pm 1.28$\tablenotemark{h}   & Galaxy group \\
5S2                              & \hmsd{15}{00}{44}{87} $\pm$ 0.17& \dmsd{78}{18}{39}{0} $\pm$ 0.5& NC        & \nodata                  & \nodata                              & \nodata &\nodata & None\\
5S3/8S1                      & \hmsd{15}{01}{11}{70} $\pm$ 0.17& \dmsd{78}{15}{20}{2} $\pm$ 0.2& NC        &  \nodata                 & \nodata                              & \nodata& \nodata& Galaxy\\
5S4                              & \hmsd{15}{01}{16}{19} $\pm$ 0.17& \dmsd{78}{12}{45}{9} $\pm$ 0.2& \nodata &  $< 11.2$              & $< 2.27 \times 10^{-4}$  & \nodata &\nodata & Galaxy\\
5S5/8S2                      & \hmsd{15}{01}{22}{67} $\pm$ 0.17& \dmsd{78}{18}{05}{7} $\pm$ 0.2& X58       & $38.5 \pm 6.4$    & $ 7.79 \times 10^{-4}$     & $0.52 \pm 0.23$ & $0.40 \pm 0.86$      & None\\
5S6/8S3                      & \hmsd{15}{02}{51}{40} $\pm$ 0.17& \dmsd{78}{18}{12}{5} $\pm$ 0.2& \nodata & $< 3.2$                 & $6.48 \times 10^{-5}$      &\nodata &\nodata & None \\
5S7/8S4                      & \hmsd{15}{03}{24}{83} $\pm$ 0.23& \dmsd{78}{17}{37}{6} $\pm$ 0.8& \nodata & $< 10.9$               & $< 2.20 \times 10^{-4}$  &\nodata &\nodata  & None \\
5S8                              & \hmsd{15}{03}{28}{07} $\pm$ 0.49& \dmsd{78}{09}{21}{5} $\pm$ 1.7& NC         & \nodata                  &  \nodata                             &\nodata& \nodata & Galaxy? \\
5T1 (RT\,19840502) & \hmsd{15}{02}{24}{61} $\pm$ 0.35& \dmsd{78}{16}{10}{1} $\pm$ 0.8& \nodata & $< 9.9$                  &  $< 2.00 \times 10^{-4}$ &\nodata& \nodata & None \\
5T2 (RT\,19840613) & \hmsd{15}{01}{38}{07} $\pm$ 0.35& \dmsd{78}{18}{40}{8} $\pm$ 0.6& \nodata & $< 12.4$                & $< 2.51 \times 10^{-4}$  &\nodata& \nodata & $z = 0.04$ galaxy\\
5T3 (RT\,19860115) & \hmsd{15}{02}{26}{40} $\pm$ 0.73& \dmsd{78}{17}{32}{4} $\pm$ 2.6& \nodata & $< 5.4$                   & $< 1.09 \times 10^{-4}$ &\nodata&  \nodata & None\\
5T4 (RT\,19860122) & \hmsd{15}{00}{50}{15} $\pm$ 0.56& \dmsd{78}{15}{39}{4} $\pm$ 2.3& \nodata & $< 9.9$                   & $< 2.00 \times 10^{-4}$ &\nodata& \nodata & None \\
5T5 (RT\,19920826) & \hmsd{15}{02}{59}{89} $\pm$ 0.58& \dmsd{78}{16}{10}{8} $\pm$ 4.0& \nodata & $< 5.8$                   & $< 1.17 \times 10^{-4}$ &\nodata& \nodata & Galaxy? \\
5T6 (RT\,19970528) & \hmsd{15}{00}{23}{55} $\pm$ 0.17& \dmsd{78}{13}{01}{4} $\pm$ 0.3& \nodata & $<13.2$                  & $< 2.67 \times 10^{-4}$ &\nodata&  \nodata & $z = 0.25$ galaxy?\\
5T7 (RT\,19990504) & \hmsd{14}{59}{46}{42} $\pm$ 0.92& \dmsd{78}{20}{29}{0} $\pm$ 1.2& X19\tablenotemark{i}       &  $40.5 \pm 7.2$     & $8.20 \times 10^{-4}$     &$-0.20 \pm 0.20$ & $5.57 \pm 3.17$       & None\tablenotemark{i} \\
8T1 (RT\,19970205) & \hmsd{15}{01}{29}{35} $\pm$ 0.17& \dmsd{78}{19}{49}{2} $\pm$ 0.2& \nodata & $< 8.7$                   & $< 1.76 \times 10^{-4}$ & \nodata & \nodata & None \\
5L1 (RT\,19870422) & \hmsd{15}{00}{50}{01} $\pm$ 0.58& \dmsd{78}{09}{45}{5} $\pm$ 1.6& \nodata & $<21.2$                  & $< 4.29 \times 10^{-4}$ &\nodata & \nodata & $z = 0.25$ galaxy\\
8L1 (RT\,20010331) & \hmsd{15}{03}{46}{18} $\pm$ 0.17& \dmsd{78}{15}{41}{7} $\pm$ 0.2& X124     & $8.8 \pm 3.6$        & $1.78 \times 10^{-4}$     & $0.38 \pm 0.44$ & $1.70 \pm 1.09$ & None \\
\enddata
\tablenotetext{a}{90\%\ position uncertainties computed by multiplying the uncertainties from B07 by 1.65.}
\tablenotetext{b}{See Table~\ref{tab:xray}. Sources marked ``NC'' are outside the coverage of the \chandra\ image or fall in the gaps between the chips.}
\tablenotetext{c} {ACIS counts are reported for \fullband\ for the four sources with cross-matches in the \chandra\ catalog (Table~\ref{tab:xray}) noted in the column headed ``\chandra\ ID''. For these sources, we tabulate the corresponding count rate from Table~\ref{tab:xray}, and the offset between the X-ray and radio positions. For sources with no match in the \chandra\ catalog, we tabulate counts in apertures centered at the radio positions, and corresponding 90\%-confidence upper limits to the \fullband\ rate. Aperture radii were 5\arcsec, except for 5S7 and 5T6, where a 7\farcs5-radius aperture was used, and 5L1, where a 10\arcsec-radius aperture was used (these three sources are rather far off-axis, where the \chandra\ PSF is somewhat elongated).}
\tablenotetext{d}{In cases where upper limits are shown, these are 90\%\ confidence Poisson limits computed following \citet{poisson}.}
\tablenotetext{e}{See Table~\ref{tab:xray}.}
\tablenotetext{f}{The uncertainty in the offset is computed by adding the 90\%\ positional uncertainties of the X-ray and radio positions in quadrature.}
\tablenotetext{g}{Redshifts are from \citet{bower}. The two transients with IDs with question marks were considered by B07 as possible counterparts, but we argue here that they are in fact too far away from the OIR sources to be likely to be associated.}
\tablenotetext{h}{The X-ray source is extended, and the offset is computed from the centroid of the X-ray position to the centroid of the radio position.}
\tablenotetext{i}{Although B07 and O10 consider the radio source to be unidentified, the nearby X-ray source is associated with a $z = 1.29$ quasar (\S~\ref{sec:x19sed}). The radio source may or may not be physically associated with the X-ray and optical / IR source (\S~\ref{sec:x19prog}).}
\end{deluxetable}

B07 also constructed a deep image of the field using all data at each frequency. The resulting images at 5 and 8.4\,GHz had an rms of 2.6 and 2.8\,\ujy, respectively. Sources which were seen in many of the individual epochs and also in the deep field image were termed ``steady'' by B07. They saw 8 steady sources at 5\,GHz, shown in Table~\ref{tab:radio} with identifiers beginning ``5S'', and 4 of these correspond to steady sources in the deep image at 8.4\,GHz, shown with labels beginning ``8S'' in Table~\ref{tab:radio}.

B07 also compared images made by making images from two months' worth of data at a time, and found two additional ``long-timescale'' transients, one at each frequency, shown with identifiers beginning ``5L'' and ``8L'' in Table~\ref{tab:radio}.

Optical and infra-red (OIR) observations of the field were also reported by B07, both with LRIS on Keck I, to limiting magnitudes of $g \approx 27.6$\,mag and $R \approx 26.6$\,mag, and with PAIRITEL to $J \approx 19.2$\,mag, $H \approx 18.5$\,mag and $\ks \approx 18.0$\,mag. Of the two long-duration transients, one has a match with a $z = 0.25$ galaxy. Of the eight short-duration transients, one has a likely match (with a $z = 0.04$ galaxy). O10 also present observations in \ks-band using the Hale 5.08-m telescope to a limiting magnitude of \ks\ ranging from $19.2$ to $20.4$ for those sources undetected in OIR by B07. They found no additional identifications. Two transients are referred to by B07 as ``possible'' matches, one with a galaxy at $z = 0.25$, and the other with a galaxy with unknown redshift (see Table~\ref{tab:radio}), but the images of O10 suggest that both of these associations are unlikely.  We consider these two, like the remaining six B07 radio transients, to be unidentified in all the OIR images from B07 and O10.  

As discussed above, B07 suggest a range of possible progenitors for these transients, and O10 expand on this discussion. O10 also suggest that the transients seen at the Nasu Pulsar Observatory by \citet{matsumura:09} and other papers from the same group may be members of the same class. Recently, \citet{paperi}, \citet{paperii}, and \citet{bower:11} suggested that the Nasu transients are probably not astronomical events, and that their quoted rate is at odds with other measurements, although the parameters of the Nasu survey are not clear from the literature \citep[see][]{ofek:11}. However, the B07 transient rate is consistent with measurements from other surveys (see, for example, \citealt{paperii}, Fig.~12), and two of the B07 transients have clear OIR counterparts. O10 show that for most classes of object, in particular extragalactic events including supernovae and gamma ray bursts, one would expect a large fraction of the transients to have associated host galaxies. They argue that the most attractive explanation is that the transients are associated with isolated old Galactic neutron stars. In order to test this hypothesis, and the flare star scenario discussed by B07, we obtained X-ray observations of the field.

\subsection{X-Ray Observations}

X-ray observations were obtained with \chandra\ during Cycle 11, using the ACIS-I instrument with no grating or filter. The exposure time was 49,441\,s and the observation start date was UT 2010 July 3 01:14:32. The observation was centered at \hmstd{15}{01}{50}{58}, \dmstd{78}{16}{50}{90} so as to encompass the majority of the VLA field. 

For our analysis, we started with the ``level 2''  data products, and these
were processed at the \chandra\ X-ray Center with pipeline
(``ASCDS'') version 8.3.  We performed further processing with the
\chandra\ Interactive Analysis of Observations (CIAO) version 4.3
software and version 4.4.1 of the Calibration Data Base (CALDB).
For the work described below, we used an image from the four
ACIS-I CCDs, including photons in the \fullband\ bandpass.

A catalog was created from the full-band image, by running the CIAO source detection algorithm ``wavdetect''. 
We used 4 different pixel binnings (1, 2, 4, and 8) in order to detect sources that are either truly extended or
sources that appear extended because the PSF broadens off-axis. We combined the catalogs from these 4 runs. A total of 128 sources were detected, and their properties are shown in Table~\ref{tab:xray}. In each run, the threshold was set such that 1 spurious source would be expected, so one would expect up to 4 of the sources in Table~\ref{tab:xray} to be spurious. In most cases, the parameters shown in Table~\ref{tab:xray} correspond to the lowest pixel binning \label{sec:binning}in which a source was detected. In some cases (such as the extended source X29, or off-axis sources), the lowest binning clearly did not enclose
all of the counts and a higher binning was used. The binning corresponding to the reported source parameters is shown in Table~\ref{tab:xray}.

\begin{deluxetable}{lllllllll}
\tablewidth{0pt}
\tabletypesize{\scriptsize}
\tablecaption{\label{tab:xray} \chandra\ sources}
\tablehead {
\colhead{\chandra\ ID} & 
\colhead{\chandra\ RA} &
\colhead{\chandra\ Dec} &
\colhead{\chandra\ pos.} &
\colhead{Bin\tablenotemark{b}} &
\colhead{ACIS counts\tablenotemark{c}} &
\colhead{Hardness} &
\colhead{VLA}\\
 & 
\colhead{(J2000)} &
\colhead{(J2000)} &
\colhead{unc.~(arcsec)\tablenotemark{a}} &
&
\colhead{(\fullband)} &
\colhead{constraint\tablenotemark{d}} &
\colhead{ID\tablenotemark{e}}&
}
\startdata
X1 & \hmsd{14}{57}{57}{57} & \dmsd{78}{15}{29}{7} & 1.46 & 4 & $45.4 \pm 9.5$ & $<0.71$ & \\
X2 & \hmsd{14}{58}{14}{94} & \dmsd{78}{16}{38}{1} & 1.63 & 4 & $60.0 \pm 12.2$ & $<0.89$ & \\
X3 & \hmsd{14}{58}{31}{24} & \dmsd{78}{18}{14}{5} & 1.25 & 2 & $28.6 \pm 6.8$ & $<0.04$ & \\
X4 & \hmsd{14}{58}{33}{49} & \dmsd{78}{15}{03}{8} & 1.29 & 2 & $45.4 \pm 8.9$ & $-0.29 \pm 0.22$ & \\
X5 & \hmsd{14}{58}{45}{15} & \dmsd{78}{15}{29}{3} & 0.95 & 1 & $8.0 \pm 3.2$ & $<0.62$ & \\
X6 & \hmsd{14}{58}{46}{87} & \dmsd{78}{18}{60}{0} & 1.98 & 4 & $23.9 \pm 7.8$ & \nodata & \\
X7 & \hmsd{14}{58}{48}{87} & \dmsd{78}{15}{13}{5} & 0.86 & 1 & $7.9 \pm 3.0$ & $<0.26$ & \\
X8 & \hmsd{14}{59}{02}{72} & \dmsd{78}{16}{58}{6} & 1.22 & 3 & $22.8 \pm 6.4$ & $0.03 \pm 0.40$ & \\
X9 & \hmsd{14}{59}{11}{82} & \dmsd{78}{20}{50}{5} & 1.11 & 3 & $36.1 \pm 8.1$ & $-0.33 \pm 0.32$ & \\
X10 & \hmsd{14}{59}{20}{11} & \dmsd{78}{17}{05}{8} & 0.85 & 1 & $8.1 \pm 3.2$ & $<0.07$ & \\
X11 & \hmsd{14}{59}{23}{81} & \dmsd{78}{18}{08}{5} & 1.02 & 2 & $9.8 \pm 3.7$ & $<0.31$ & \\
X12 & \hmsd{14}{59}{25}{15} & \dmsd{78}{15}{20}{7} & 1.40 & 3 & $30.4 \pm 8.0$ & $0.24 \pm 0.41$ & \\
X13 & \hmsd{14}{59}{26}{92} & \dmsd{78}{14}{28}{5} & 1.37 & 3 & $38.6 \pm 8.0$ & $0.36 \pm 0.29$ & \\
X14 & \hmsd{14}{59}{26}{98} & \dmsd{78}{16}{49}{8} & 0.93 & 1 & $25.2 \pm 6.1$ & $0.09 \pm 0.25$ & \\
X15 & \hmsd{14}{59}{30}{92} & \dmsd{78}{14}{05}{9} & 1.28 & 3 & $22.8 \pm 6.6$ & $<0.22$ & \\
X16 & \hmsd{14}{59}{32}{13} & \dmsd{78}{11}{20}{2} & 1.69 & 4 & $24.8 \pm 8.1$ & \nodata & \\
X17 & \hmsd{14}{59}{35}{61} & \dmsd{78}{12}{59}{4} & 0.93 & 2 & $23.7 \pm 5.8$ & $-0.60 \pm 0.31$ & \\
X18 & \hmsd{14}{59}{45}{84} & \dmsd{78}{21}{09}{6} & 1.01 & 2 & $16.6 \pm 4.8$ & $0.32 \pm 0.43$ & \\
X19 & \hmsd{14}{59}{46}{99} & \dmsd{78}{20}{23}{7} & 0.83 & 2 & $40.5 \pm 7.2$ & $-0.20 \pm 0.20$ & 5T7\\
X20 & \hmsd{14}{59}{50}{62} & \dmsd{78}{10}{46}{7} & 1.43 & 3 & $12.9 \pm 5.0$ & $0.14 \pm 0.75$ & \\
X21 & \hmsd{14}{59}{55}{38} & \dmsd{78}{11}{16}{7} & 1.16 & 3 & $24.8 \pm 6.9$ & $0.30 \pm 0.41$ & \\
X22 & \hmsd{14}{59}{57}{59} & \dmsd{78}{13}{49}{5} & 0.72 & 1 & $9.6 \pm 3.5$ & $-0.28 \pm 0.51$ & \\
X23 & \hmsd{14}{59}{57}{89} & \dmsd{78}{21}{27}{6} & 0.83 & 1 & $26.8 \pm 6.2$ & $-0.36 \pm 0.25$ & \\
X24 & \hmsd{15}{00}{00}{57} & \dmsd{78}{22}{53}{2} & 0.81 & 1 & $8.4 \pm 3.3$ & $<0.31$ & \\
X25 & \hmsd{15}{00}{02}{85} & \dmsd{78}{17}{41}{3} & 1.02 & 2 & $15.5 \pm 4.6$ & $>-0.03$ & \\
X26 & \hmsd{15}{00}{05}{30} & \dmsd{78}{14}{47}{3} & 0.70 & 1 & $7.5 \pm 2.8$ & $<0.21$ & \\
X27 & \hmsd{15}{00}{06}{04} & \dmsd{78}{18}{16}{1} & 0.74 & 1 & $6.6 \pm 2.8$ & $-0.21 \pm 0.54$ & \\
X28 & \hmsd{15}{00}{06}{45} & \dmsd{78}{18}{35}{1} & 0.69 & 1 & $48.2 \pm 7.7$ & $-0.07 \pm 0.16$ & \\
X29 & \hmsd{15}{00}{08}{62} & \dmsd{78}{12}{48}{6} & 1.16 & 4 & $166.8 \pm 21.1$ & $-0.62 \pm 0.11$ & 5S1\\
X30 & \hmsd{15}{00}{09}{28} & \dmsd{78}{16}{18}{9} & 0.82 & 1 & $18.1 \pm 4.7$ & $0.40 \pm 0.33$ & \\
X31 & \hmsd{15}{00}{12}{56} & \dmsd{78}{21}{03}{4} & 1.11 & 1 & $16.1 \pm 5.0$ & $-0.45 \pm 0.46$ & \\
X32 & \hmsd{15}{00}{13}{00} & \dmsd{78}{15}{45}{8} & 1.04 & 2 & $16.4 \pm 4.9$ & $>0.24$ & \\
X33 & \hmsd{15}{00}{15}{41} & \dmsd{78}{24}{20}{9} & 1.38 & 2 & $18.7 \pm 5.9$ & $-0.38 \pm 0.59$ & \\
X34 & \hmsd{15}{00}{19}{57} & \dmsd{78}{14}{04}{8} & 1.00 & 2 & $12.5 \pm 4.1$ & $-0.49 \pm 0.48$ & \\
X35 & \hmsd{15}{00}{24}{61} & \dmsd{78}{13}{41}{5} & 0.71 & 1 & $30.0 \pm 5.8$ & $-0.30 \pm 0.22$ & \\
X36 & \hmsd{15}{00}{25}{27} & \dmsd{78}{15}{49}{5} & 0.77 & 1 & $8.5 \pm 3.2$ & $>-0.07$ & \\
X37 & \hmsd{15}{00}{26}{05} & \dmsd{78}{15}{18}{1} & 0.73 & 1 & $23.5 \pm 5.1$ & $-0.39 \pm 0.31$ & \\
X38 & \hmsd{15}{00}{27}{14} & \dmsd{78}{11}{29}{0} & 1.43 & 3 & $15.1 \pm 5.4$ & $-0.02 \pm 0.62$ & \\
X39 & \hmsd{15}{00}{29}{41} & \dmsd{78}{13}{02}{9} & 0.76 & 1 & $12.6 \pm 4.0$ & $-0.31 \pm 0.31$ & \\
X40 & \hmsd{15}{00}{32}{58} & \dmsd{78}{21}{12}{0} & 0.90 & 2 & $9.6 \pm 3.6$ & $>-0.54$ & \\
X41 & \hmsd{15}{00}{33}{99} & \dmsd{78}{14}{58}{5} & 0.76 & 1 & $6.3 \pm 2.6$ & $<0.59$ & \\
X42 & \hmsd{15}{00}{35}{55} & \dmsd{78}{21}{36}{6} & 0.66 & 1 & $332.7 \pm 19.9$ & $-0.37 \pm 0.06$ & \\
X43 & \hmsd{15}{00}{43}{05} & \dmsd{78}{09}{38}{2} & 1.06 & 2 & $16.5 \pm 4.9$ & $-0.48 \pm 0.47$ & \\
X44 & \hmsd{15}{00}{44}{19} & \dmsd{78}{26}{41}{8} & 1.32 & 4 & $103.1 \pm 14.3$ & $0.05 \pm 0.19$ & \\
X45 & \hmsd{15}{00}{45}{54} & \dmsd{78}{12}{12}{1} & 0.71 & 1 & $4.7 \pm 2.2$ & \nodata & \\
X46 & \hmsd{15}{00}{47}{01} & \dmsd{78}{21}{20}{2} & 0.86 & 2 & $43.5 \pm 7.7$ & $0.09 \pm 0.18$ & \\
X47 & \hmsd{15}{00}{48}{72} & \dmsd{78}{18}{00}{3} & 1.12 & 3 & $19.4 \pm 6.2$ & $>0.16$ & \\
X48 & \hmsd{15}{00}{52}{64} & \dmsd{78}{09}{00}{6} & 0.85 & 3 & $65.2 \pm 9.3$ & $-0.38 \pm 0.18$ & \\
X49 & \hmsd{15}{00}{54}{07} & \dmsd{78}{14}{17}{9} & 0.97 & 2 & $3.7 \pm 2.2$ & \nodata & \\
X50 & \hmsd{15}{00}{54}{39} & \dmsd{78}{08}{13}{2} & 1.31 & 3 & $14.9 \pm 5.2$ & $<0.02$ & \\
X51 & \hmsd{15}{00}{55}{19} & \dmsd{78}{13}{48}{3} & 0.72 & 1 & $26.2 \pm 5.6$ & $-0.45 \pm 0.27$ & \\
X52 & \hmsd{15}{00}{55}{43} & \dmsd{78}{17}{56}{8} & 0.67 & 1 & $4.5 \pm 2.2$ & $<0.26$ & \\
X53 & \hmsd{15}{00}{58}{19} & \dmsd{78}{15}{11}{7} & 1.55 & 4 & $36.0 \pm 10.6$ & $0.23 \pm 0.49$ & \\
X54 & \hmsd{15}{01}{05}{65} & \dmsd{78}{12}{56}{3} & 0.72 & 1 & $21.0 \pm 4.8$ & $0.13 \pm 0.34$ & \\
X55 & \hmsd{15}{01}{05}{82} & \dmsd{78}{15}{39}{6} & 0.76 & 1 & $6.0 \pm 2.6$ & $<-0.06$ & \\
X56 & \hmsd{15}{01}{15}{61} & \dmsd{78}{23}{02}{7} & 0.81 & 1 & $35.8 \pm 7.2$ & $-0.07 \pm 0.19$ & \\
X57 & \hmsd{15}{01}{20}{47} & \dmsd{78}{20}{29}{1} & 0.80 & 1 & $7.6 \pm 3.0$ & $<0.05$ & \\
X58 & \hmsd{15}{01}{22}{67} & \dmsd{78}{18}{06}{1} & 0.68 & 1 & $38.5 \pm 6.4$ & $0.52 \pm 0.23$ & 5S5/8S2\\
X59 & \hmsd{15}{01}{23}{82} & \dmsd{78}{09}{55}{6} & 0.91 & 2 & $16.9 \pm 4.8$ & $<-0.23$ & \\
X60 & \hmsd{15}{01}{23}{88} & \dmsd{78}{06}{47}{9} & 0.73 & 2 & $165.5 \pm 15.1$ & $-0.39 \pm 0.09$ & \\
X61 & \hmsd{15}{01}{26}{36} & \dmsd{78}{24}{39}{6} & 0.90 & 1 & $33.3 \pm 7.5$ & $-0.16 \pm 0.20$ & \\
X62 & \hmsd{15}{01}{26}{86} & \dmsd{78}{07}{38}{9} & 1.37 & 3 & $14.7 \pm 5.5$ & $0.09 \pm 0.45$ & \\
X63 & \hmsd{15}{01}{29}{75} & \dmsd{78}{21}{19}{3} & 0.72 & 1 & $17.0 \pm 4.5$ & $-0.08 \pm 0.40$ & \\
X64 & \hmsd{15}{01}{29}{75} & \dmsd{78}{19}{14}{0} & 0.73 & 1 & $7.0 \pm 2.8$ & $0.27 \pm 0.58$ & \\
X65 & \hmsd{15}{01}{30}{50} & \dmsd{78}{25}{11}{9} & 0.78 & 2 & $109.2 \pm 13.2$ & $-0.41 \pm 0.13$ & \\
X66 & \hmsd{15}{01}{31}{14} & \dmsd{78}{20}{18}{4} & 0.75 & 1 & $14.1 \pm 4.0$ & $0.34 \pm 0.55$ & \\
X67 & \hmsd{15}{01}{32}{91} & \dmsd{78}{12}{46}{6} & 1.00 & 2 & $13.0 \pm 4.2$ & $<-0.10$ & \\
X68 & \hmsd{15}{01}{35}{97} & \dmsd{78}{06}{53}{2} & 1.04 & 3 & $32.2 \pm 7.7$ & $-0.32 \pm 0.32$ & \\
X69 & \hmsd{15}{01}{37}{58} & \dmsd{78}{10}{03}{2} & 0.76 & 1 & $6.2 \pm 2.6$ & \nodata & \\
X70 & \hmsd{15}{01}{41}{14} & \dmsd{78}{09}{38}{5} & 1.28 & 3 & $12.5 \pm 4.8$ & $<0.74$ & \\
X71 & \hmsd{15}{01}{44}{79} & \dmsd{78}{15}{26}{6} & 0.64 & 1 & $172.6 \pm 13.3$ & $-0.25 \pm 0.09$ & \\
X72 & \hmsd{15}{01}{45}{39} & \dmsd{78}{08}{36}{8} & 0.77 & 1 & $80.1 \pm 10.5$ & $-0.61 \pm 0.14$ & \\
X73 & \hmsd{15}{01}{45}{95} & \dmsd{78}{12}{03}{9} & 0.73 & 1 & $24.3 \pm 5.2$ & $0.53 \pm 0.33$ & \\
X74 & \hmsd{15}{01}{48}{39} & \dmsd{78}{15}{41}{0} & 1.04 & 2 & $7.0 \pm 3.0$ & \nodata & \\
X75 & \hmsd{15}{01}{53}{42} & \dmsd{78}{05}{34}{5} & 1.64 & 4 & $18.4 \pm 5.7$ & $>-0.27$ & \\
X76 & \hmsd{15}{01}{54}{60} & \dmsd{78}{09}{08}{5} & 1.06 & 1 & $11.4 \pm 3.9$ & $<0.37$ & \\
X77 & \hmsd{15}{01}{54}{69} & \dmsd{78}{26}{13}{5} & 1.83 & 4 & $22.1 \pm 7.6$ & $<0.59$ & \\
X78 & \hmsd{15}{01}{56}{25} & \dmsd{78}{18}{59}{5} & 1.72 & 4 & $24.0 \pm 9.0$ & $<0.48$ & \\
X79 & \hmsd{15}{01}{59}{30} & \dmsd{78}{10}{35}{0} & 0.68 & 1 & $6.7 \pm 2.6$ & \nodata & \\
X80 & \hmsd{15}{02}{02}{31} & \dmsd{78}{15}{22}{1} & 0.66 & 1 & $41.9 \pm 6.7$ & $-0.01 \pm 0.18$ & \\
X81 & \hmsd{15}{02}{04}{83} & \dmsd{78}{08}{47}{5} & 0.78 & 2 & $5.2 \pm 2.6$ & \nodata & \\
X82 & \hmsd{15}{02}{06}{85} & \dmsd{78}{05}{16}{9} & 2.28 & 4 & $33.1 \pm 8.4$ & $>0.04$ & \\
X83 & \hmsd{15}{02}{07}{00} & \dmsd{78}{16}{22}{7} & 0.66 & 1 & $46.2 \pm 6.9$ & $-0.32 \pm 0.21$ & \\
X84 & \hmsd{15}{02}{07}{50} & \dmsd{78}{15}{48}{7} & 0.68 & 1 & $19.7 \pm 4.6$ & $-0.33 \pm 0.33$ & \\
X85 & \hmsd{15}{02}{08}{80} & \dmsd{78}{16}{32}{3} & 0.72 & 1 & $8.7 \pm 3.2$ & $-0.11 \pm 0.49$ & \\
X86 & \hmsd{15}{02}{12}{95} & \dmsd{78}{09}{08}{6} & 1.37 & 3 & $13.1 \pm 4.7$ & \nodata & \\
X87 & \hmsd{15}{02}{14}{95} & \dmsd{78}{18}{30}{6} & 0.68 & 1 & $5.4 \pm 2.4$ & $<0.59$ & \\
X88 & \hmsd{15}{02}{15}{93} & \dmsd{78}{11}{38}{1} & 0.73 & 1 & $11.2 \pm 3.6$ & $-0.26 \pm 0.39$ & \\
X89 & \hmsd{15}{02}{17}{58} & \dmsd{78}{08}{36}{9} & 0.89 & 1 & $9.9 \pm 3.6$ & $<-0.26$ & \\
X90 & \hmsd{15}{02}{19}{21} & \dmsd{78}{14}{47}{7} & 0.72 & 1 & $14.6 \pm 4.0$ & $-0.02 \pm 0.41$ & \\
X91 & \hmsd{15}{02}{21}{11} & \dmsd{78}{24}{34}{5} & 1.53 & 4 & $44.1 \pm 11.3$ & $>0.08$ & \\
X92 & \hmsd{15}{02}{24}{68} & \dmsd{78}{16}{52}{0} & 0.72 & 1 & $14.5 \pm 4.0$ & $0.11 \pm 0.42$ & \\
X93 & \hmsd{15}{02}{24}{92} & \dmsd{78}{13}{48}{9} & 1.34 & 2 & $13.5 \pm 4.4$ & $>0.16$ & \\
X94 & \hmsd{15}{02}{27}{06} & \dmsd{78}{08}{42}{9} & 0.92 & 2 & $8.5 \pm 3.3$ & $<0.40$ & \\
X95 & \hmsd{15}{02}{35}{73} & \dmsd{78}{23}{48}{0} & 0.89 & 1 & $32.7 \pm 7.0$ & $-0.38 \pm 0.22$ & \\
X96 & \hmsd{15}{02}{35}{97} & \dmsd{78}{10}{51}{2} & 1.04 & 2 & $11.7 \pm 4.1$ & $-0.26 \pm 0.53$ & \\
X97 & \hmsd{15}{02}{37}{60} & \dmsd{78}{16}{27}{0} & 0.67 & 1 & $3.7 \pm 2.0$ & \nodata & \\
X98 & \hmsd{15}{02}{38}{98} & \dmsd{78}{09}{02}{3} & 0.96 & 1 & $20.5 \pm 5.7$ & $-0.18 \pm 0.30$ & \\
X99 & \hmsd{15}{02}{41}{57} & \dmsd{78}{23}{43}{5} & 1.02 & 3 & $24.8 \pm 6.4$ & $-0.35 \pm 0.34$ & \\
X100 & \hmsd{15}{02}{44}{86} & \dmsd{78}{20}{04}{3} & 0.68 & 1 & $5.4 \pm 2.4$ & $<0.26$ & \\
X101 & \hmsd{15}{02}{46}{47} & \dmsd{78}{17}{56}{7} & 0.69 & 1 & $11.9 \pm 3.6$ & $-0.03 \pm 0.53$ & \\
X102 & \hmsd{15}{02}{47}{44} & \dmsd{78}{12}{36}{0} & 1.02 & 2 & $22.9 \pm 5.8$ & $0.25 \pm 0.33$ & \\
X103 & \hmsd{15}{02}{48}{22} & \dmsd{78}{07}{59}{0} & 1.58 & 4 & $24.2 \pm 8.2$ & $>-0.11$ & \\
X104 & \hmsd{15}{02}{51}{38} & \dmsd{78}{17}{35}{6} & 0.70 & 1 & $32.1 \pm 5.9$ & $-0.01 \pm 0.23$ & \\
X105 & \hmsd{15}{02}{51}{80} & \dmsd{78}{11}{27}{0} & 0.81 & 1 & $10.9 \pm 3.6$ & $-0.04 \pm 0.62$ & \\
X106 & \hmsd{15}{02}{53}{37} & \dmsd{78}{09}{30}{5} & 0.96 & 2 & $19.3 \pm 5.3$ & $0.21 \pm 0.39$ & \\
X107 & \hmsd{15}{02}{56}{01} & \dmsd{78}{10}{21}{6} & 1.25 & 3 & $26.7 \pm 6.9$ & $0.32 \pm 0.36$ & \\
X108 & \hmsd{15}{03}{02}{99} & \dmsd{78}{18}{20}{6} & 0.70 & 1 & $26.2 \pm 5.4$ & $-0.12 \pm 0.27$ & \\
X109 & \hmsd{15}{03}{03}{61} & \dmsd{78}{14}{51}{9} & 0.76 & 1 & $9.2 \pm 3.3$ & $-0.07 \pm 0.36$ & \\
X110 & \hmsd{15}{03}{04}{48} & \dmsd{78}{22}{13}{0} & 1.55 & 4 & $21.9 \pm 7.5$ & $<0.41$ & \\
X111 & \hmsd{15}{03}{07}{02} & \dmsd{78}{17}{48}{9} & 0.71 & 1 & $20.7 \pm 4.9$ & $-0.22 \pm 0.26$ & \\
X112 & \hmsd{15}{03}{10}{12} & \dmsd{78}{21}{36}{5} & 1.88 & 4 & $31.2 \pm 8.9$ & $>-0.45$ & \\
X113 & \hmsd{15}{03}{11}{45} & \dmsd{78}{18}{48}{5} & 0.74 & 1 & $9.7 \pm 3.5$ & $<-0.20$ & \\
X114 & \hmsd{15}{03}{13}{38} & \dmsd{78}{17}{47}{0} & 1.22 & 2 & $13.8 \pm 4.4$ & $0.28 \pm 0.43$ & \\
X115 & \hmsd{15}{03}{14}{18} & \dmsd{78}{22}{51}{3} & 1.44 & 3 & $39.1 \pm 7.9$ & $<0.43$ & \\
X116 & \hmsd{15}{03}{16}{02} & \dmsd{78}{10}{04}{1} & 1.86 & 4 & $29.0 \pm 9.6$ & $>0.53$ & \\
X117 & \hmsd{15}{03}{16}{31} & \dmsd{78}{13}{02}{8} & 0.78 & 1 & $17.9 \pm 4.6$ & $-0.43 \pm 0.34$ & \\
X118 & \hmsd{15}{03}{23}{14} & \dmsd{78}{14}{28}{4} & 0.68 & 1 & $97.6 \pm 10.6$ & $-0.22 \pm 0.11$ & \\
X119 & \hmsd{15}{03}{24}{06} & \dmsd{78}{18}{08}{4} & 1.34 & 4 & $21.7 \pm 8.9$ & $0.38 \pm 0.52$ & \\
X120 & \hmsd{15}{03}{25}{76} & \dmsd{78}{14}{39}{9} & 0.75 & 1 & $13.8 \pm 4.1$ & $-0.29 \pm 0.43$ & \\
X121 & \hmsd{15}{03}{29}{45} & \dmsd{78}{16}{06}{9} & 0.81 & 1 & $16.3 \pm 4.4$ & $<-0.20$ & \\
X122 & \hmsd{15}{03}{37}{93} & \dmsd{78}{18}{14}{9} & 0.95 & 2 & $11.4 \pm 3.9$ & $0.38 \pm 0.44$ & \\
X123 & \hmsd{15}{03}{42}{10} & \dmsd{78}{17}{08}{3} & 0.72 & 1 & $15.2 \pm 4.2$ & $0.07 \pm 0.29$ & \\
X124 & \hmsd{15}{03}{45}{73} & \dmsd{78}{15}{40}{7} & 0.95 & 2 & $8.8 \pm 3.6$ & $0.38 \pm 0.44$ & 8L1\\
X125 & \hmsd{15}{04}{07}{84} & \dmsd{78}{13}{16}{4} & 1.50 & 4 & $39.9 \pm 10.3$ & $>-0.11$ & \\
X126 & \hmsd{15}{04}{40}{84} & \dmsd{78}{15}{33}{8} & 1.96 & 4 & $40.2 \pm 10.4$ & \nodata & \\
X127 & \hmsd{15}{04}{56}{27} & \dmsd{78}{16}{42}{8} & 1.53 & 4 & $37.9 \pm 9.8$ & $>0.31$ & \\
X128 & \hmsd{15}{05}{13}{24} & \dmsd{78}{18}{17}{9} & 1.53 & 4 & $41.1 \pm 9.2$ & $>0.03$ & \\
\enddata
\tablenotetext{a}{90\%\ positional uncertainty radius, including the $0\farcs64$ systematic \chandra\ pointing uncertainty added in quadrature.}
\tablenotetext{b}{Binning in which the source was best detected (\S~\ref{sec:binning})}
\tablenotetext{c}{Background subtracted}
\tablenotetext{d}{Based on the fact that the hardness must lie in the range -1 to 1}
\tablenotetext{e}{See Table~\ref{tab:radio}.}
\end{deluxetable}

\clearpage

Event filtering was performed using $3 \times 3$ pixel islands. This approach, compared to the use of $5 \times 5$ pixel islands, may result in a slightly larger fraction of cosmic rays being misidentified as real sources, but also has the benefit of being less likely to reject real astronomical sources.\label{sec:islands}

\begin{figure}[t]
\centering
\includegraphics[width=0.45\linewidth,draft=false]{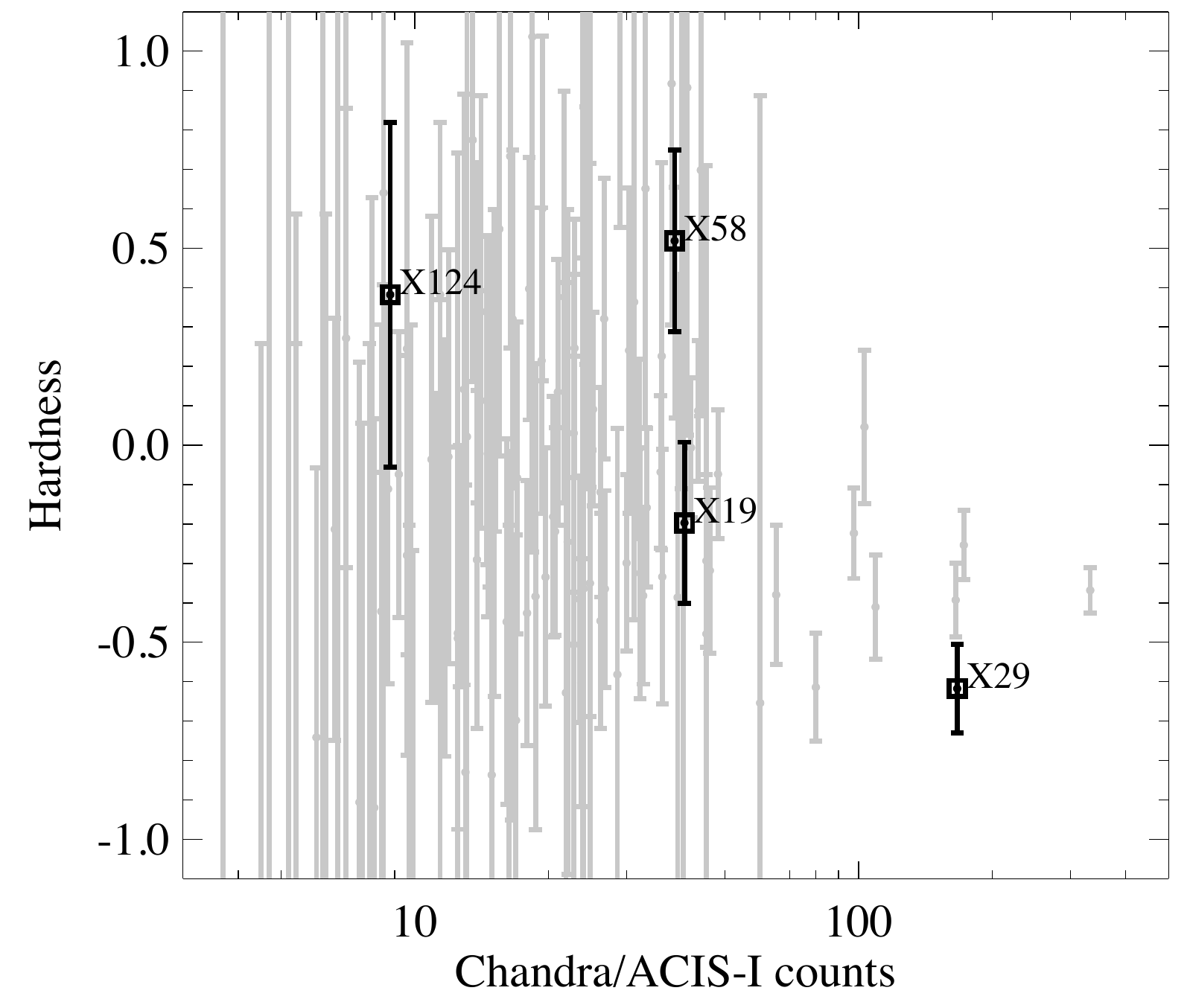}
\caption{\label{fig:hr}
Hardness ratio as a function of number of \fullband\ counts for the sources in Table~\ref{tab:xray}. The four cases where there are possible matches to our VLA sources (Table~\ref{tab:radio}) are marked.
}
\end{figure}

Hardness ratios were computed as $HR = ((C_2-C_1)/(C_1+C_2))$ where $C_1$ is the counts (after background subtraction) in the $0.3 - 2$\,keV band and $C_2$ the counts in $2 - 10$\,keV. We tabulate $HR$ in Table~\ref{tab:xray}. In some cases the formal errors on $HR$ resulted in values outside the range $-1 < HR < 1$; here we truncate the values at these limits and report upper or lower limits on $HR$ as appropriate. In Fig.~\ref{fig:hr}, we plot $HR$ as a function of number of counts for the sources in Table~\ref{tab:xray}.

\begin{figure}
\centering
\includegraphics[width=\linewidth,draft=false]{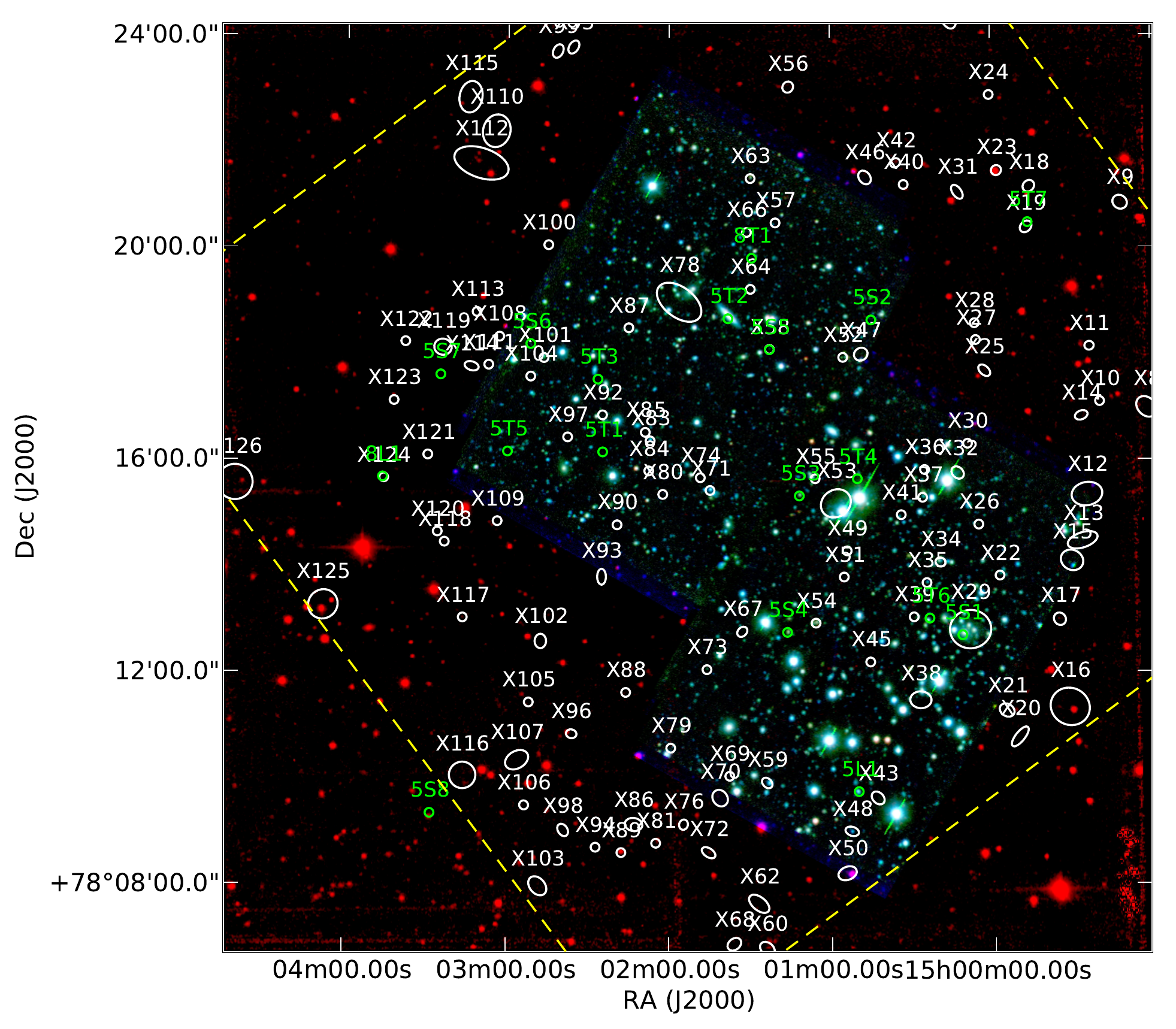}
\caption{\label{fig:optxr}
Optical / IR image of the field with radio and X-ray positions overlaid. The color image was made from the data from B07: Keck $g$- and $R$-band images (in the blue and green channels respectively) and the mean of the PAIRITEL $J$-, $H$-, and \ks-band images (in the red channel). Green circles are the radio sources from B07, as labelled in Table~\ref{tab:radio}, and white circles are the \chandra\ sources reported here, as labelled in Table~\ref{tab:xray}. The sizes of the white ellipses correspond to the size of the \chandra\ sources as determined by the wavdetect software, except for cases where the size was 5\arcsec\ or smaller, in which case we show a circle with a radius of 5\arcsec\ for clarity. The green circles are plotted with a radius of 5\arcsec\ for clarity; typical radio position uncertainties are $\sim 1$\arcsec\ (Table~\ref{tab:radio}). The yellow dashed box shows the edge of the \chandra\ field.
}
\end{figure}

In Fig.~\ref{fig:optxr}, we show a multi-wavelength view of the field, created by combining the $g$-, $R$-, $J$-, $H$- and \ks-band data from B07, and overlaying the B07 radio catalog (Table~\ref{tab:radio}), and the X-ray catalog from Table~\ref{tab:xray}.

\section{Multi-Frequency Matching}

\subsection{Radio -- X-Ray Catalog Matching}

\begin{figure}
\centering
\includegraphics[width=0.45\linewidth,draft=false]{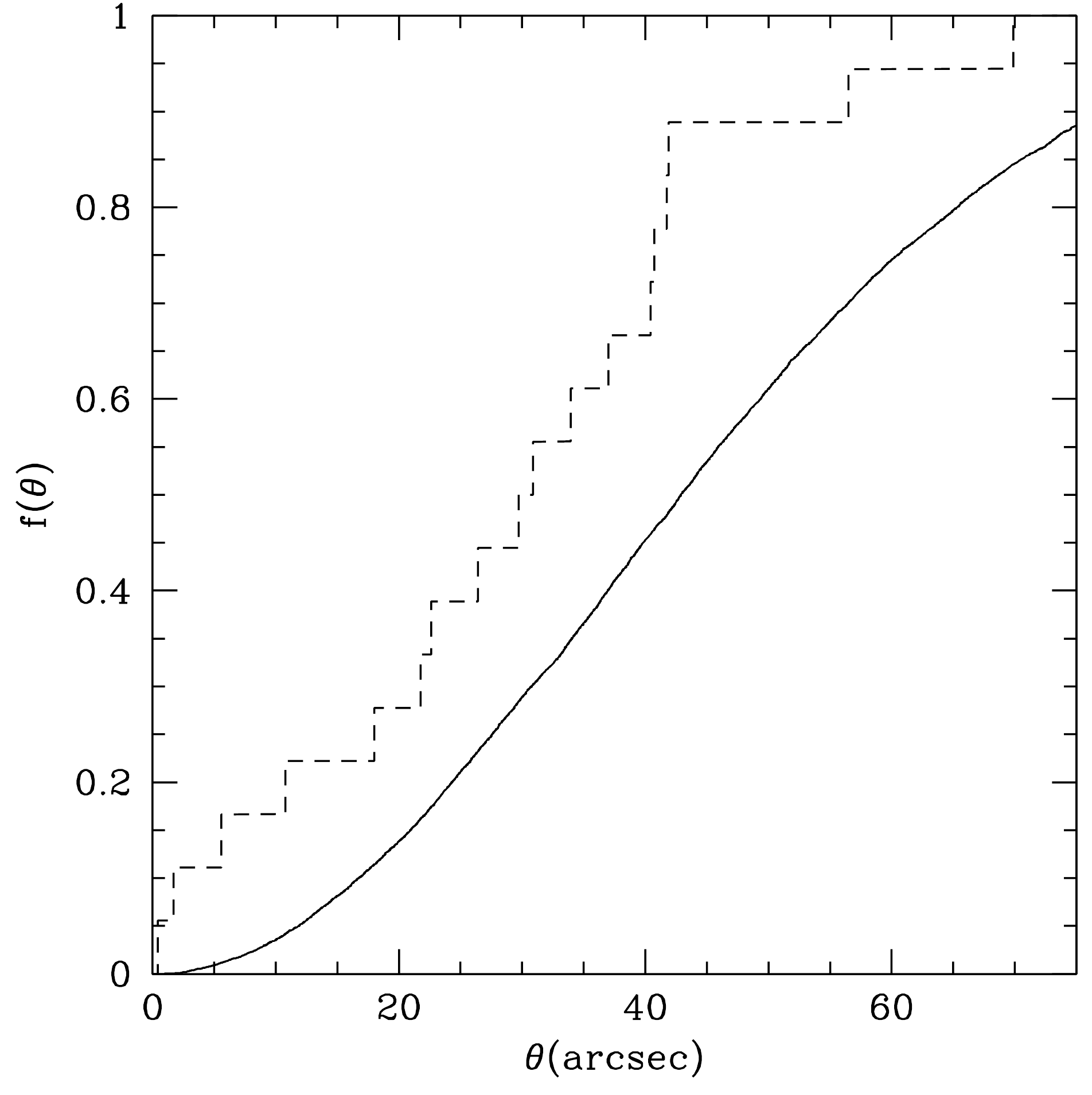}\hspace{0.05\linewidth}%
\includegraphics[width=0.45\linewidth,draft=false]{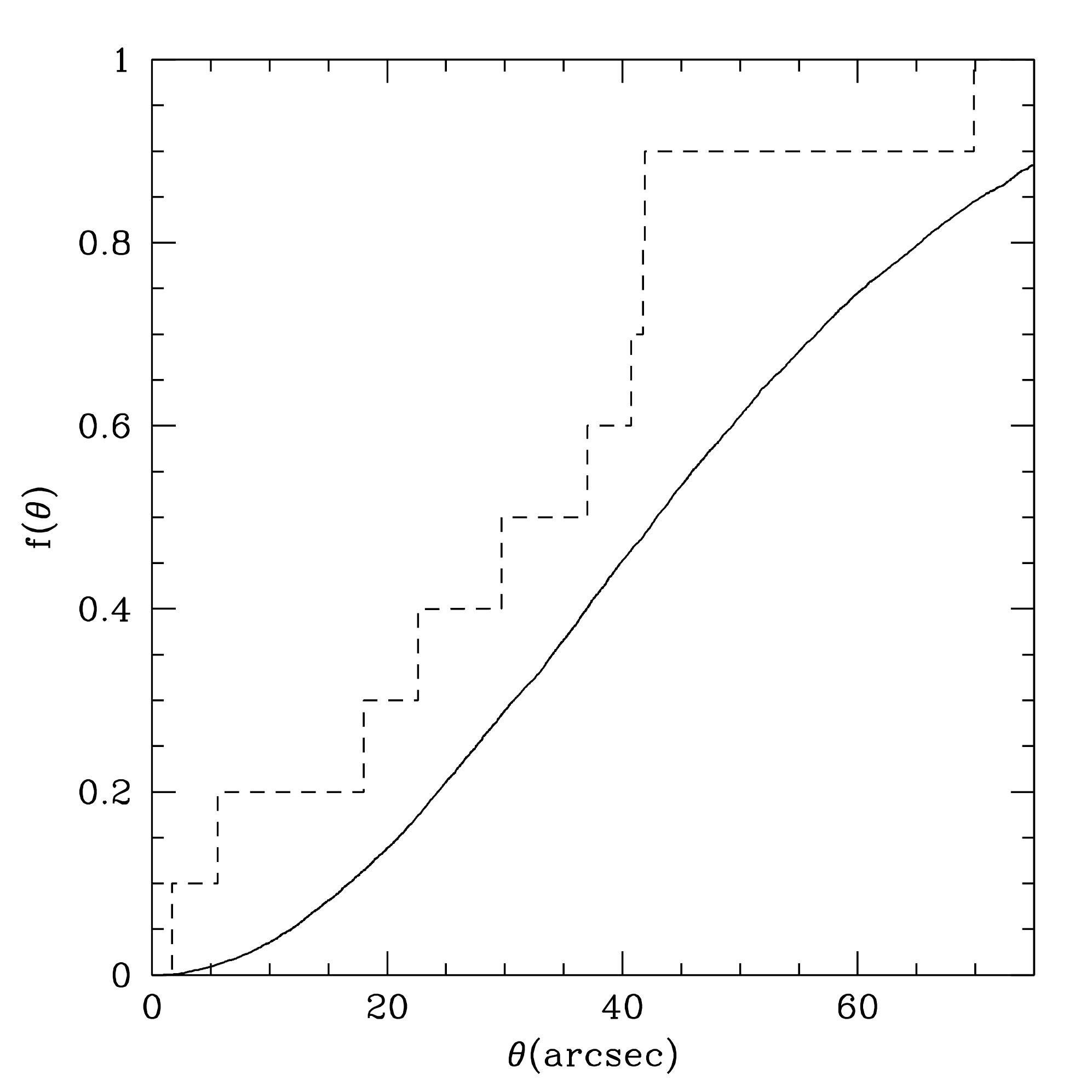}
\caption{\label{fig:matchfrac}
{\em Left:} Fraction of sources, $f(\theta)$, with a match at radius $\theta_m \leq \theta$, shown as a function of $\theta$, where we matched the 18 sources in the B07 VLA catalog with our \chandra\ catalog (dashed line) and with 1000 Monte Carlo iterations of a mock catalog with the same areal density as our \chandra\ catalog (solid line). The curve from the Monte Carlo simulations is a very good match to the expected analytic form, $f_{mock}(\theta) = 1 - e^{-\pi \rho \theta^2}$, where $\rho = 1.2 \times 10^{-4}$\,arcsec$^{-2}$ is the areal density of \chandra\ sources. The real \chandra\ catalog shows an excess above random of matches to the VLA sources, showing that at least some of the radio sources are really associated with X-ray sources.
{\em Right:} The same plot, but removing the 8 steady VLA sources and plotting the curves only for the 10 transient sources. An excess above random is still seen.
}
\end{figure}

We searched around the radio positions (Table~\ref{tab:radio}) for matches in the X-ray catalog (Table~\ref{tab:xray}). In order to determine the maximum radius where a match is believable, we ran Monte Carlo simulations. The areal density of \chandra\ sources in our image is $\sim 1600\,{\rm deg}^{-2}$. We ran 1000 iterations of a routine where we created mock X-ray catalogs with the same density (over an area twice as large as our VLA field), and then searched for matches in this simulated catalog around the VLA positions. We calculated $f(\theta)$, the fraction of VLA sources with a match in the catalog closer than radius $\theta$, for both the mock and real \chandra\ catalogs. At small $\theta$, we expect few matches in the mock catalog, but as $\theta$ increases, the fraction of sources with matches also increases due to chance associations with neighboring sources. In Fig.~\ref{fig:matchfrac}, we plot $f(\theta)$ for both the mock and real catalogs. For the mock catalog, the match fraction is very closely approximated by the expected analytic form,
\begin{equation}
\label{eqn:thetamock} 
f_{mock}(\theta) = 1 - e^{-\pi \rho \theta^2},
\end{equation}
where $\rho = 1.2 \times 10^{-4}$\,arcsec$^{-2}$ is the areal density of \chandra\ sources. For a match at a given radius $\theta$, $f_{mock}(\theta)$ gives an estimate of the probability that the match is due to chance. This assumes a uniform surface density of X-ray sources; if sources in the mock catalogs were clustered (as they are in the real catalog), $f_{mock}(\theta)$ would tend to be lower over areas of the X-ray image away from concentrations of X-ray sources, leading to a tendency to overestimate the false match probability. Proper motion is also not accounted for; we discuss this further in Section~\ref{sec:proper1}.

For $\theta = 11\farcs7$, $f_{mock}(\theta) = 0.05$, \ie, 5\%\ of VLA sources have a match in the mock catalogs within this radius. If sources in the X-ray catalog are physically associated with sources in the radio catalog, we expect an excess of matches at small $\theta$ over the expectation from the Monte Carlo simulations, and indeed this is seen when we compare $f_{mock}(\theta)$ and $f_{real}(\theta)$ in Fig.~\ref{fig:matchfrac}. The 5\%\ false match probability leads us to expect $\lesssim 1$ false match out of our 18 radio sources at a radius of 11\farcs7, but for matches that are closer than this, the probability increases that matched sources are physically associated rather than superposed by chance. 

We used Equation~\ref{eqn:thetamock} to calculate values of $f_{mock}(\theta)$ (\ie, the probability that the association is due to chance) for radio sources with a match in the real \chandra\ catalog closer than $11\farcs7$. Two of the steady sources, and two of the radio transients (one short, and one long) have possible \chandra\ counterparts, and we discuss each of these cases below (Sections~\ref{sec:matchstead} and \ref{sec:matchtran}).

Bandwidth smearing in the radio image is not a significant issue. At worst it increases the size of unresolved sources close to the edge of the field in the 5\,GHz VLA A-array observations of B07 by around 70\%, but in most cases is significantly less than this. Even in the worst case it should not make a large difference to the statistical significance of matches between the X-ray and radio catalogs.

Due to the small number of matches between X-ray and radio sources, it is not possible to look for systematic offsets between the X-ray and radio astrometric frames. However, VLA astrometric accuracy is typically $\sim 0\farcs1$ or better, so systematic errors are likely dominated by the $0\farcs64$ \chandra\ pointing uncertainty, which is already included in the positional uncertainties reported in Table~\ref{tab:xray}. We show in \S~\ref{sec:icrs} that the \chandra\ data are well-matched to the optical data. Sub-arcsecond systematic offsets would have only a small effect on the confidence levels for the matches discussed below.

\subsubsection{Two X-Ray Matches to Steady Radio Sources\label{sec:matchstead}}

Steady radio source 5S1 was matched to X-ray source X29 at a distance of $10\farcs79$. The 90\%\ positional uncertainty of X29 is 1\farcs16. The positional uncertainties reported by B07 for source 5S1 evaluate to a 90\%\ uncertainty of  0\farcs53. Adding these in quadrature gives a total 90\%\ positional uncertainty of $1\farcs28$. Source X29 is extended, and 5S1 lies within the boundaries of the \chandra\ source (Fig.~\ref{fig:optxr}). From Equation~\ref{eqn:thetamock}, $f(\theta_{mock} = 10\farcs79 \pm 1\farcs28) =  0.0429^{+0.0105}_{-0.0094}$, so such an association has a $4.29^{+1.05}_{-0.94}$\% probability of happening by chance. There are 18 sources altogether (eight steady and ten transient), and the binomial probability of a false association for one or more of them is 54.6\%, so the formal confidence for this association is just 45.4\%. However, source X29 is bright and extended, and is clearly associated with a group of galaxies in the OIR image (Fig.~\ref{fig:optxr}), one of which is clearly the host of the radio source, so the sources both do appear to have a physical association with the galaxy group (\S~\ref{sec:5s1}).

Steady source 5S5/8S2 was matched to source X58 at a distance of $0\farcs40$. The 90\%\ positional uncertainty of X58 is $0\farcs68$. The 90\%\ positional uncertainty of 5S5 is $0\farcs53$. Adding these in quadrature gives a total 90\%\ positional uncertainty of $0\farcs86$, so the match is well within the uncertainties. The false match probability is $0.006^{+0.037}_{-0.006}$\%, or $0.1$\%\ for at least one out of 18 sources. So we can be $99.9$\%\ confident that the association is real.

\begin{figure}
\centering
\includegraphics[width=0.45\linewidth,draft=false]{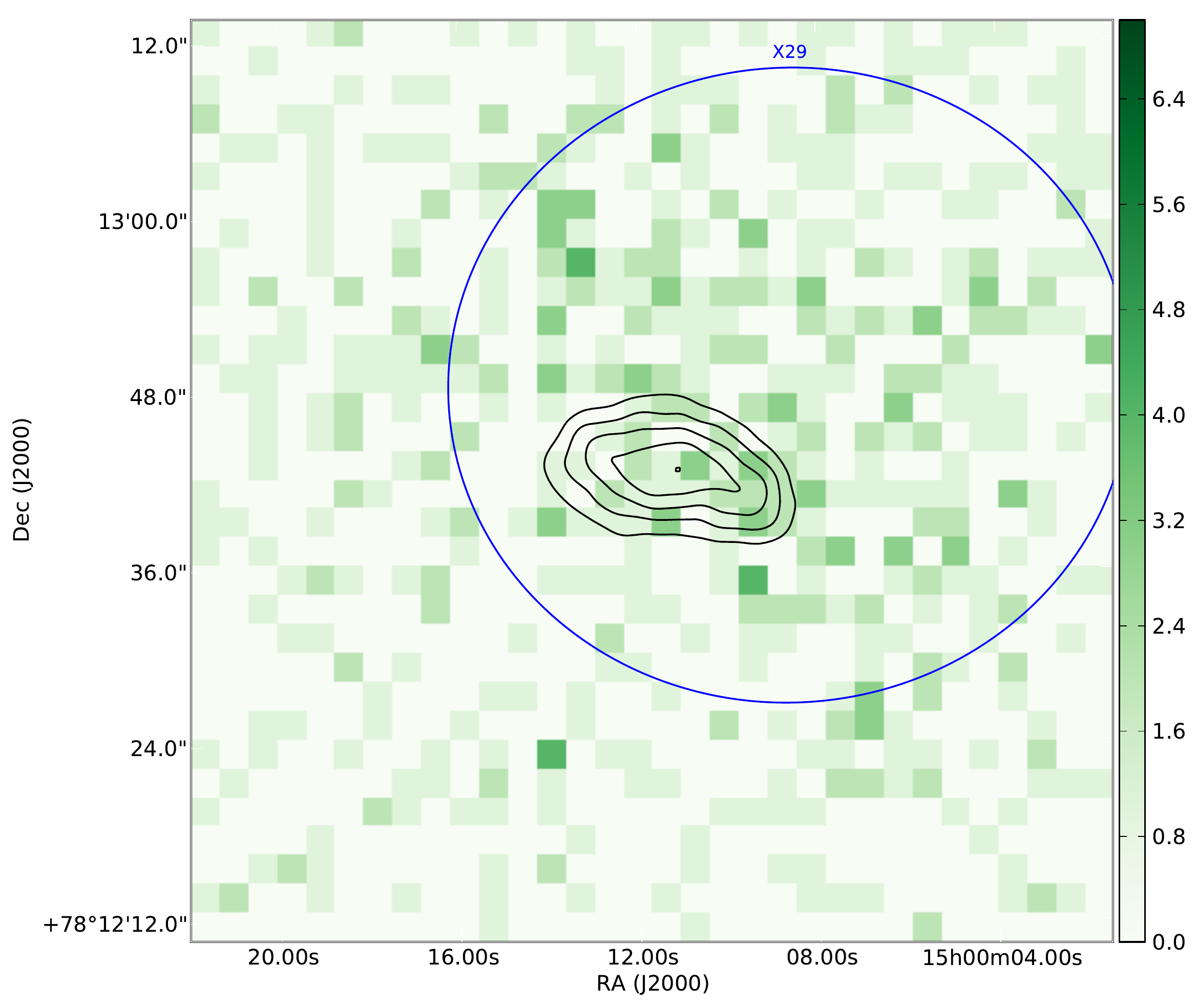}%
\includegraphics[width=0.45\linewidth,draft=false]{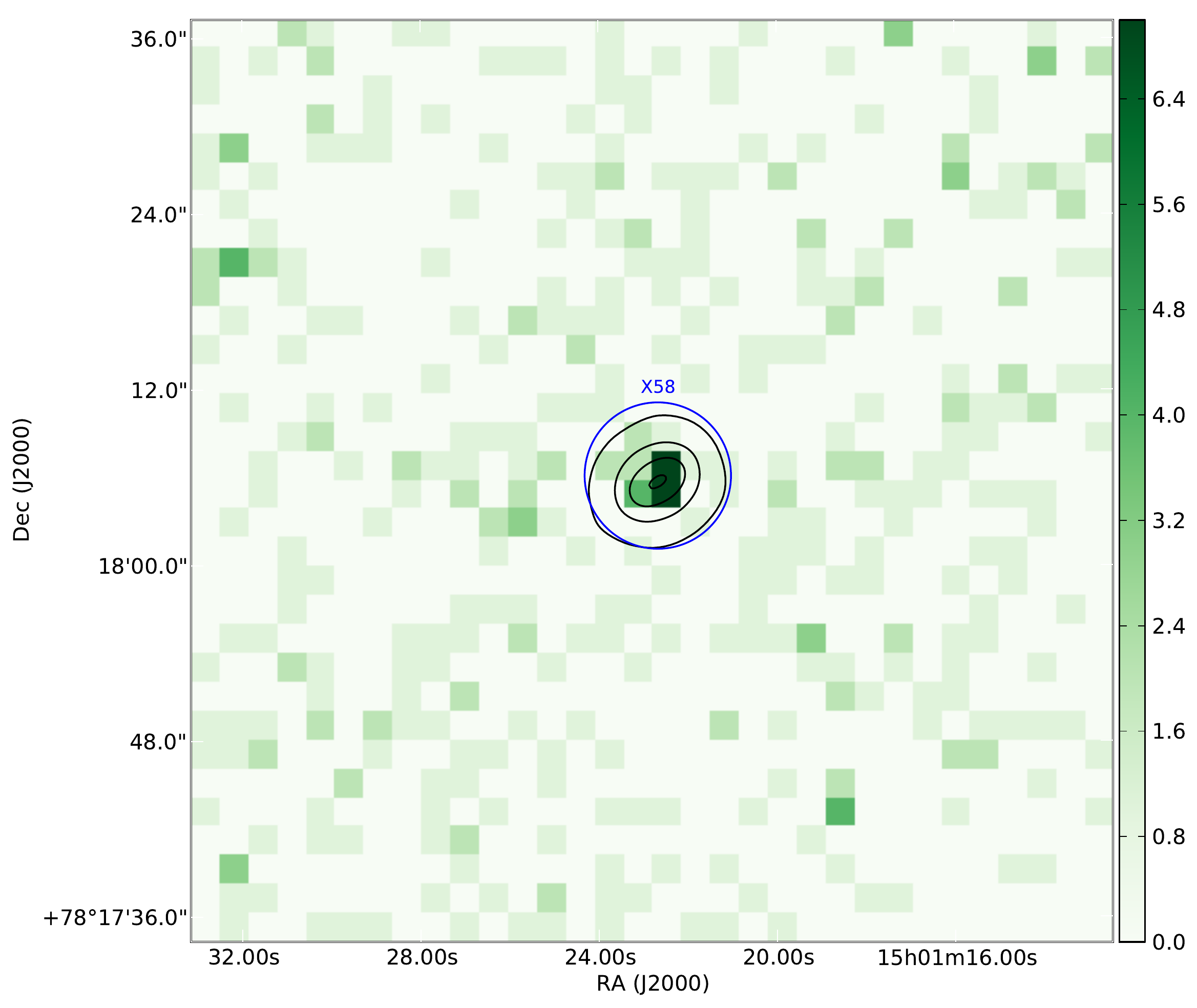}
\caption{\label{fig:postages}
Images of the two fields with X-ray counterparts to our steady radio sources 5S1 (left) and 5S5 (right). The colors in the image represent the number of \chandra\ \fullband\ counts per pixel. The image is binned in blocks of $4 \times 4$ pixels. The radio data are from the deep 5\,GHz image, with contours from 15 -- 35\,\ujypbm\ in steps of 5\,\ujypbm\ for 5S1, and from 50 -- 200\,\ujypbm\ in steps of 50\,\ujypbm\ for 5S5.
The size of the blue ellipse for source X29 corresponds to the size of the \chandra\ source as determined by the wavdetect software. For source X58 the fitted size was $<5$\arcsec\ and the circle is shown with a radius of 5\arcsec\ for clarity.
}
\end{figure}

In Fig.~\ref{fig:postages} we show X-ray images with radio contours overlaid of the two fields with X-ray counterparts to our steady sources.

\subsubsection{Two Possible X-Ray Matches to Transient Radio Sources\label{sec:matchtran}}

Long (duration $\sim 2$ months) transient source 8L1 was matched to source X124 at a distance of $1\farcs70$. The 90\%\ position uncertainty for X124 is 0\farcs95. The radio position uncertainties reported by B07 for source 8L1 evaluate to a 90\%\ uncertainty of $0\farcs54$. Adding these in quadrature gives a total 90\%\ positional uncertainty of 1\farcs28. So the measured positional offset is just a little larger than the formal positional uncertainty. From Equation~\ref{eqn:thetamock}, $f_{mock}(\theta = 1\farcs70 \pm 1\farcs28) = 0.0011^{+0.0022}_{-0.0010}$, so such an association has a $0.11^{+0.22}_{-0.10}$\%\ probability of happening by chance. There are ten radio transients, and the binomial probability of one or more of them being associated with a \chandra\ source by chance is $1.0$\%. Hence, we can be $99$\%\ confident that the association between 8L1 and X124 is real. We note, however, that X124 was detected with $8.8 \pm 3.6$ counts, so the X-ray source is not a very strong detection. Additionally, this source is not present in a catalog filtered using $5 \times 5$ pixel islands (Section~\ref{sec:islands}). Examination of the event list before filtering shows that four out of 21 events occurred at the same time in the same pixel. These events are likely not due to photons from the source, but may be due to a cosmic ray or hot pixel. The absence of X124 from the $5 \times 5$ pixel filtered catalog means that we cannot rule out that it may be a spurious source due to cosmic rays or particle background. Additional X-ray data would be useful to confirm or reject this X-ray detection.

Short duration transient source 5T7 was matched to source X19 at a distance of $5\farcs57$. The positional uncertainty of source X19 is 0\farcs83. The positional uncertainty of source 5T7 is 3\farcs06. Adding these in quadrature for a total uncertainty of 3\farcs17, we see that the match radius is almost twice the formal uncertainties. 
However, the X-ray source is towards the edge of the X-ray field and appears somewhat extended, and the synthesized beam of the radio observations in which this transient was detected is relatively large (FWHM $28\arcsec \times 13\arcsec$); bandwidth smearing for this source is not a significant problem, though (broadening the radio source by around 10\%\ in the radial direction).
Equation~\ref{eqn:thetamock} shows that the false match probability is $1.16^{+1.68}_{-0.95}$\%. The binomial probability of one or more out of ten transients being associated with a \chandra\ source this close is 11.0\%, so we can be 89.0\%\ confident that this association is real. We discuss this source further in Section~\ref{sec:x19}.

\begin{figure}
\centering
\includegraphics[width=0.45\linewidth,draft=false]{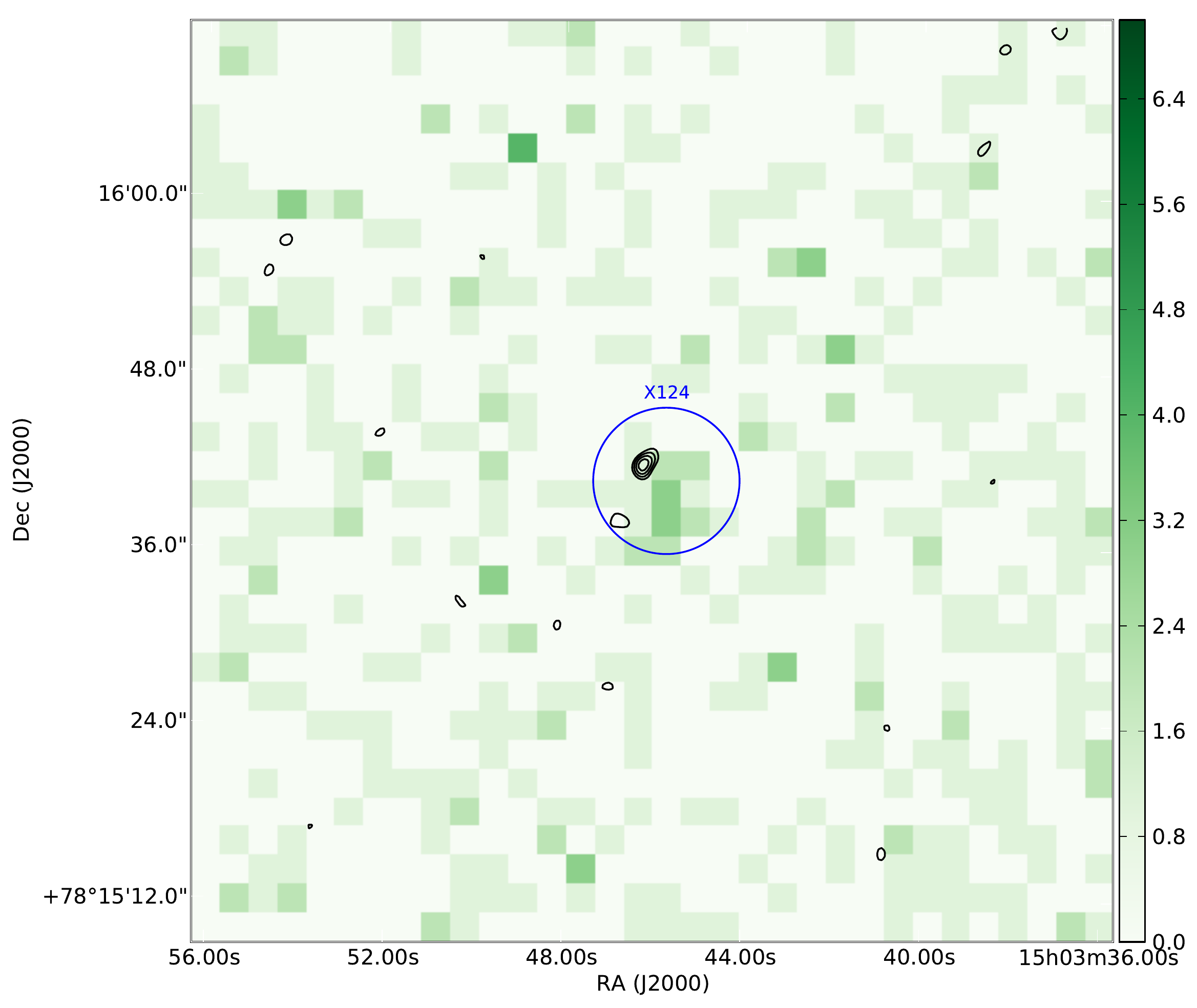}%
\includegraphics[width=0.45\linewidth,draft=false]{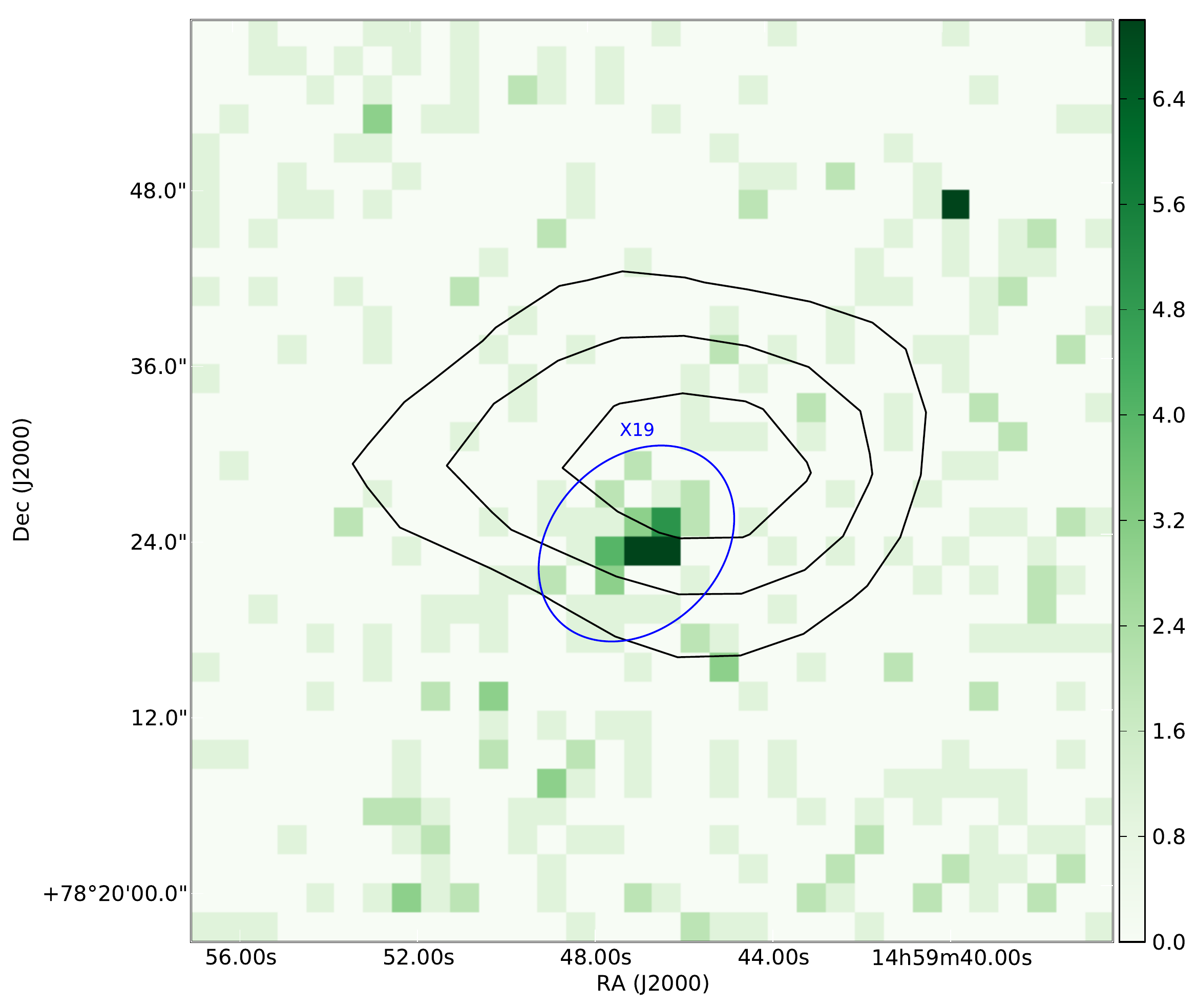}
\caption{\label{fig:postage}
Images of the two fields with possible X-ray counterparts to our radio transients 8L1 (left) and 5T7 (right). The colors in the image represent the number of \chandra\ \fullband\ counts per pixel. The image is binned in blocks of $4 \times 4$ pixels. The radio contours are from 15 -- 35\,\ujypbm\ in steps of 5\,\ujypbm\ for the two-month average VLA B-array data for 8L1, and from 150 -- 250\,\ujypbm\ in steps of 50\,\ujypbm\ for the single-epoch lower resolution VLA D-array data for 5T7.
The size of the blue ellipse for source X19 corresponds to the size of the \chandra\ source as determined by the wavdetect software. For source X124 the fitted size was $<5$\arcsec\ and the circle is shown with a radius of 5\arcsec\ for clarity.
}
\end{figure}

In Fig.~\ref{fig:postage}, we show X-ray images with radio contours overlaid of the two fields with possible X-ray counterparts to our radio transients.

\subsection{X-Ray Upper Limits at Radio Positions}

For the eight transients and three steady sources without a counterpart from our \chandra\ catalog within $11\farcs7$ (not including the three steady sources which are outside the area of our \chandra\ coverage), we measured counts from the \chandra\ \fullband\ image. We used apertures with radii ranging from 5 -- 10\arcsec\ (depending on how far away from the center of the \chandra\ image the positions were, to account for distortion of the PSF) centered at the radio positions. The counts are reported in Table~\ref{tab:radio}. We also tabulate 90\%\ Poisson upper limits for the count rates following \citet{poisson}.

\subsection{Optical / IR matches}

We generated a catalog from the $R$-band image of B07, who report an astrometric uncertainty relative to the International Celestial Reference System (ICRS) of 250 mas in each axis for this image.\label{sec:icrs} We matched this $R$-band catalog to our \chandra\ catalog, to check for systematic astrometric shifts in our \chandra\ data. For the 40 \chandra\ sources with a match in the $R$-band catalog within a radius of 10\arcsec, we found a median offset of $\Delta{\rm RA} = 0\farcs27 \pm 3\farcs91$, $\Delta{\rm Dec} = 0\farcs20 \pm 1\farcs60$. We conclude that our \chandra\ data are astrometrically well-matched to the ICRS. We neglect these small offsets in our analysis of the matches, because they are smaller than the other errors.

In Table~\ref{tab:radio}, we note the OIR matches as reported by B07 for the transient and long transient radio sources. We also report where there are OIR counterparts for the steady sources as seen in the LRIS and PAIRITEL images. Two sources (5S3 and 5S4) have clear counterparts which appear as faint, small angular size sources in the OIR images. One source (5S8) has a probable association with an IR source which is outside the footprint of the optical coverage. These three sources are likely AGNs or perhaps starburst galaxies at intermediate redshifts. None has an X-ray counterpart.

\label{sec:5s1}Source 5S1 is associated with a galaxy in a group or cluster which appears to be at intermediate redshift. The cluster is coincident with an extended X-ray source (X29) which is presumably due to bremsstrahlung from the cluster halo (\S~\ref{sec:cluster}). Transient source 5T6 is also quite nearby, and may be associated with a galaxy at $z = 0.25$, as reported by B07, although as noted above the offset between the galaxy and the radio position is rather large to be plausible. It is possible that the potential host galaxy is a member (on the outskirts) of this cluster --- it is 42\arcsec\ from the cluster radio source 5S1, and 35\arcsec\ from the centroid of the X-ray halo X29, which if the cluster is also at $z = 0.25$, correspond to 163\,kpc and 136\,kpc respectively, within the virial radius for even a poor cluster.

For the sources with OIR identifications, we searched the NASA Extragalactic Database\footnote{\url{http://nedwww.ipac.caltech.edu/}} for redshifts or other information about the counterparts, but no further information was found.

\subsection{Proper Motions\label{sec:proper1}}

For nearby sources such as brown dwarfs, high proper motions may result in positional mismatches between the same source in the radio, OIR, and X-ray data, which can complicate the interpretation of the data. For the two radio transients with nearby X-ray matches, the time between the X-ray and radio observations was $\sim 10$\,yr. From the offsets in Table~\ref{tab:radio}, this implies that if the X-ray and radio sources were exactly coincident at the time of the radio transient (and ignoring any systematic position errors), the progenitors of 8L1 and 5T7 would have proper motions (with 90\%\ confidence uncertainties) of $\mu = 0.18 \pm 0.12$~arcsec~yr$^{-1}$ and $\mu = 0.50 \pm 0.28$~arcsec~yr$^{-1}$, respectively. 

\subsection{Source X19\label{sec:x19}}

\subsubsection{Spectral Fits}

\begin{figure}
\centering
\includegraphics[width=0.75\linewidth,draft=false]{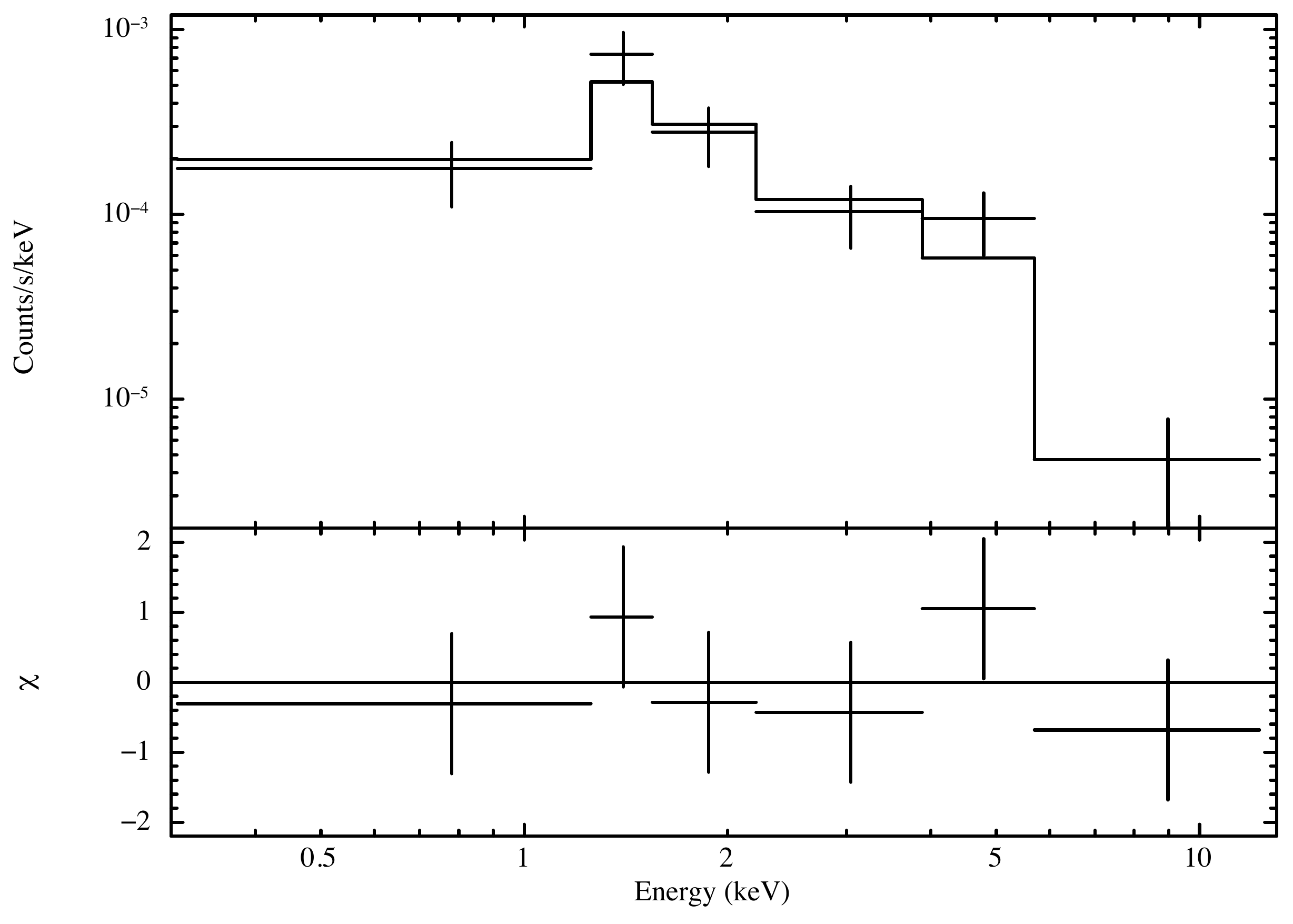}
\caption{\label{fig:x19}
{\em Top:} X-ray spectrum of source X19. The model shown is the best-fitting absorbed power-law with $N_H  < 1.2 \times 10^{22}$\,cm$^{-2}$, and $\Gamma = 1.7^{+0.9}_{-0.8}$ (Table~\ref{tab:x19}).
{\em Bottom:} The $\Delta \chi$ residuals (i.e. residuals divided by uncertainties), denoted $\chi$, for the fit to the spectrum shown in the top panel.
}
\end{figure}

Source X19 was detected with 40.5 counts in our X-ray data, and we are able to fit models to the X-ray spectrum (Fig.~\ref{fig:x19}), as opposed to simply predicting fluxes from basic models for the fainter sources. Using the XSPEC software, we fit the fluxes using three models: a power law, where the photon index and absorbing column were allowed to vary, and blackbody and bremsstrahlung models where the temperature and absorbing column were allowed to vary. Formal errors were determined using the \citet{cash} statistic. Best fit parameters with 90\%\ confidence limits, along with the reduced \chisq\ values for the best-fitting models, and associated fluxes 
are shown in Table~\ref{tab:x19}. \label{sec:specfit}

\begin{deluxetable}{lllll}
\tablewidth{0pt}
\tabletypesize{\scriptsize}
\tablecaption{\label{tab:x19} Fits for Source X19}
\tablehead {
\colhead{Model} & 
\colhead{$N_H$} &
\colhead{$kT$ or $\Gamma$} &
\colhead{Flux\tablenotemark{a}} &
\colhead{\chisq}\\
&
\colhead{ ($10^{21}$\,cm$^{-2}$)} &
&
\colhead{ ($10^{-15}$\,\esc)} &
\colhead{~(for best fit)\tablenotemark{b}} 
}
\startdata
Power Law & $4^{+8}_{-4} $ & $\Gamma = 1.7^{+0.9}_{-0.8}$ & 20 & 2.8 \\[4pt]
Blackbody & $0^{+3}_{-0} $ & $kT = 0.85^{+0.21}_{-0.17}$\,keV & 12 & 5.5 \\[4pt]
Bremsstrahlung & $3^{+6}_{-3}$ & $kT = 11^{+\infty}_{-8.2}$\,keV  & 18 & 2.9 \\
\enddata
\tablenotetext{a}{\fullband, unabsorbed, computed for the best fit parameters.}
\tablenotetext{b}{With three degrees of freedom.}
\end{deluxetable}

The X19 spectrum is about equally well fit by an absorbed power law with $\Gamma = 1.7$ as by a bremsstrahlung model with $kT = 11$\,keV. A single-temperature blackbody is a worse fit to the data than these two models. There are insufficient data to perform two-component fits to the spectrum.

\subsubsection{X-Ray Variability}

\begin{figure}
\centering
\includegraphics[width=0.45\linewidth,draft=false]{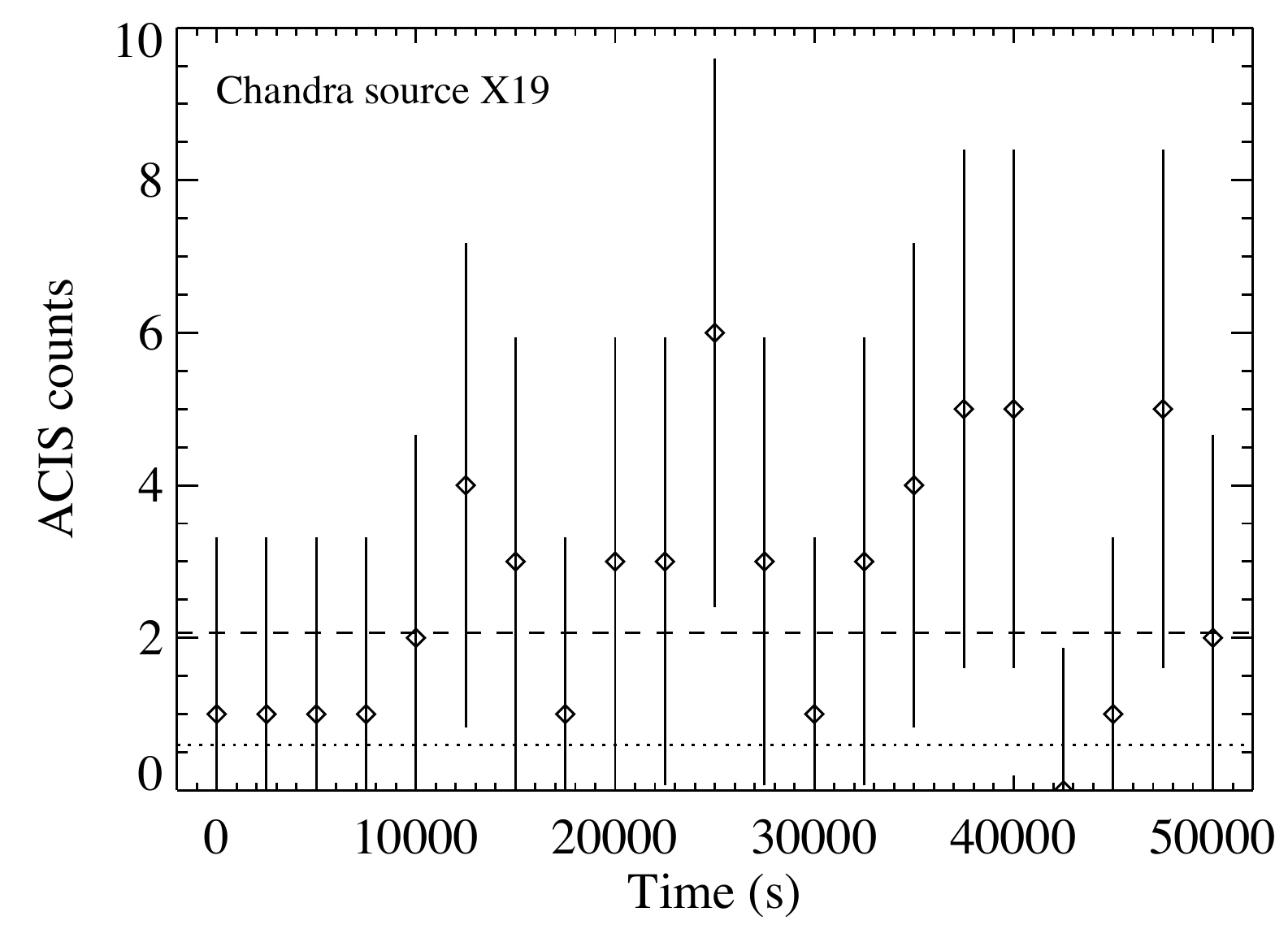}\hspace{0.05\linewidth}%
\includegraphics[width=0.45\linewidth,draft=false]{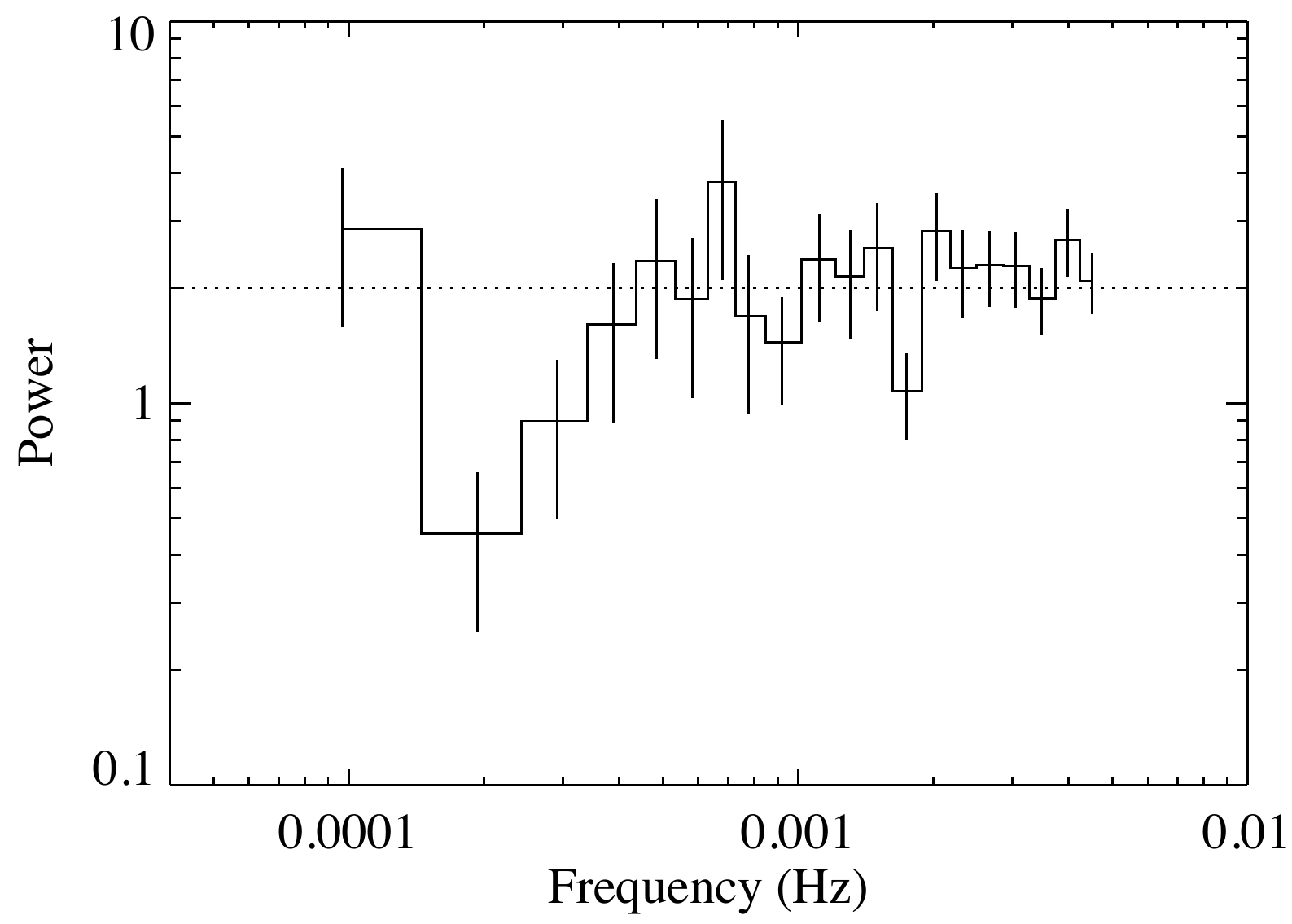}
\caption{\label{fig:x19lc}
{\em Left:} Light curve for source X19 during our \chandra\ observation. The dashed line shows the weighted average of the \chandra\ counts. The dotted line shows the background count rate.
\label{fig:x19pow}
{\em Right:} Power spectrum for source X19, using the \citet{leahy:83} normalization, where a power of 2 indicates consistency with Poisson noise with no additional variability from the source.
}
\end{figure}

In Fig.~\ref{fig:x19lc}, we show the X-ray light curve for X19.  By eye there appears to be some variability but the error bars are large, and most of them are consistent with the weighted average of the \chandra\ counts. 
A Kolmogorov-Smirnov test shows a 32\%\ chance that the source is constant, \ie, variability is detected at only the $1\sigma$ level. 

The power spectrum for X19 is shown in Fig.~\ref{fig:x19pow}. For most frequencies, the variability is consistent with the Poisson noise level with no additional variability from the source, although there is tentative evidence of variability at low frequencies. Additional epochs of X-ray data could help to determine if the source is truly variable on longer timescales. However, with only a small number of counts per second, it will remain difficult to detect variability in this source on short timescales using \chandra\ (unless it brightens significantly in the future).

\subsubsection{Multiwavelength Counterparts}

Although 5T7, the source apparently matched to X19, has no counterpart in the deep OIR imaging of B07 or O10, it is offset from the X19 position by 5\farcs57. On examination of the $K_s$-band images of the field presented in O10 (their Fig.~1, bottom right panel), we noticed a faint source offset from the radio position, but which appears to be consistent with the X-ray position. The source is not present in the $K_s$-band data of B07, so it must be fainter than the detection limit in B07 ($K_s \sim 18$) but brighter than the detection limit in O10 ($K_s \sim 19.2$). From its appearance in the O10 data, we estimate $K_s =18.5 \pm 0.3$. The source is also not visible to $H \sim 18.5$ and $J \sim 19.2$ in the B07 data.

We searched the Wide-field Infrared Survey Explorer (WISE) catalog \citep{wise} for counterparts to X19, and found that it was detected in the two shortest wavelength bands (3.4 and 4.6\,\micron, with magnitudes $17.16 \pm 0.10$ and $16.14 \pm 0.13$ respectively), and undetected at the two longer wavelengths (12 and 22\,\micron, to 12.9 and 9.1\,mag respectively).

On examination of the Second Palomar Observatory Sky Survey (POSS-II) digitized images \citep{poss2} we also found faint sources at the same position in $B$- and $R$-band, and no detection in $I$-band. Quoted limiting magnitudes are $B_j = 22.5$, $R_c = 20.8$, and $I_c = 19.5$. From the POSS-II images, we estimate $B_j = 21.0 \pm 0.5$, $R_c = 20.0 \pm 0.5$, and $I_c > 19.5$.

We also searched the Galaxy Evolution Explorer \citep[GALEX;][]{galex} Release 6 catalog for ultraviolet (UV) counterparts to X19, and discovered a source with a near-UV (1750 -- 2800\AA) magnitude in the AB system of $22.43\pm0.34$, corresponding to a Vega magnitude of $20.66$. In the far-UV (1350 -- 1750\AA), there is no GALEX detection to AB $\sim 22.3$\,mag, corresponding to a Vega magnitude fainter than 19.9.

\begin{figure}
\centering
\includegraphics[width=0.89\linewidth,draft=false]{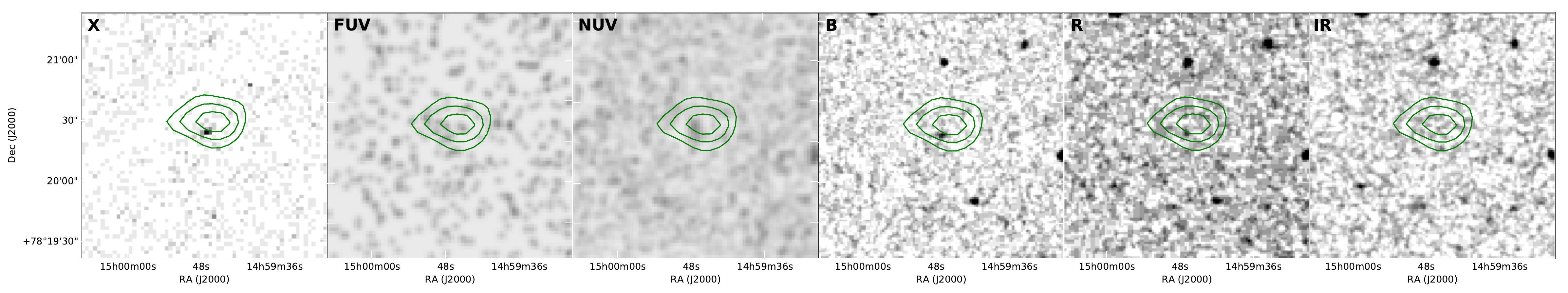}
\includegraphics[width=\linewidth,draft=false]{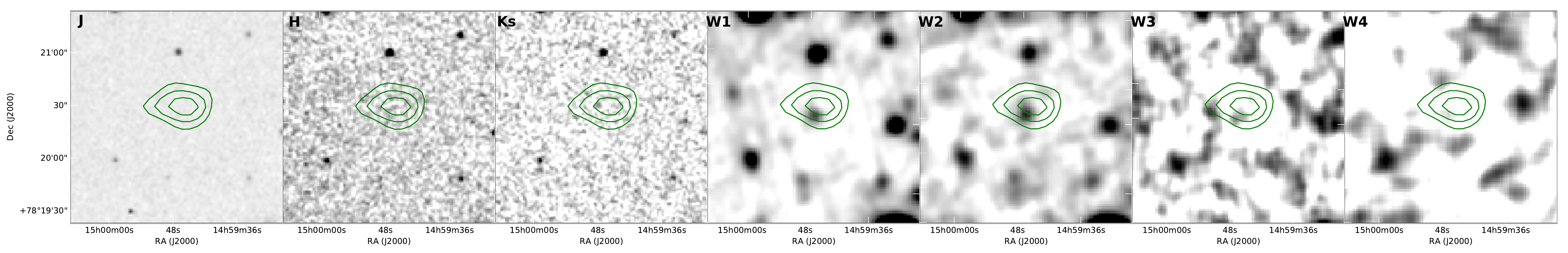}
\caption{\label{fig:x19post}
{Postage stamp images of X19. From top left to bottom right, in order of increasing wavelength: \chandra; GALEX far-UV and near-UV; POSS-II $B_j$, $R_c$, and $I_c$; $J$, $H$, and $K_s$ from B07; WISE channels 1 -- 4. Overlaid on each image are radio contours are from 150 -- 250\,\ujypbm\ in steps of 50\,\ujypbm\ for the single-epoch VLA D-array data for 5T7.
}
}
\end{figure}

Postage stamp images at the position of X19 are shown in Fig.~\ref{fig:x19post}.

\subsubsection{Proper Motion\label{sec:proper2}}

As noted in Section~\ref{sec:proper1}, if the offset between the positions of 5T7 (seen at UT 1999 May 4) and X19 (observed UT 2010 July 3) is due to movement of the source, a proper motion of $0.18 \pm 0.12$~arcsec~yr$^{-1}$ is required. However, the position of the OIR counterpart to X19 as seen in the $B_j$, $R_c$, and $I_c$-band POSS-II data (taken at UT 1994 April 6, 1994 June 09, and 1997 June 28, respectively) is coincident with the WISE position (data taken in early 2010), as well as the GALEX position (UT 2006 November 5). Therefore, proper motion cannot explain the 5\farcs57 offset between the 5T7 and X19 positions, which are bracketed by epochs where the UV/OIR counterpart to X19 is not seen to move.


\subsubsection{Spectral Energy Distribution Fitting\label{sec:x19sed}}

\begin{figure}
\centering
\includegraphics[width=\linewidth,draft=false]{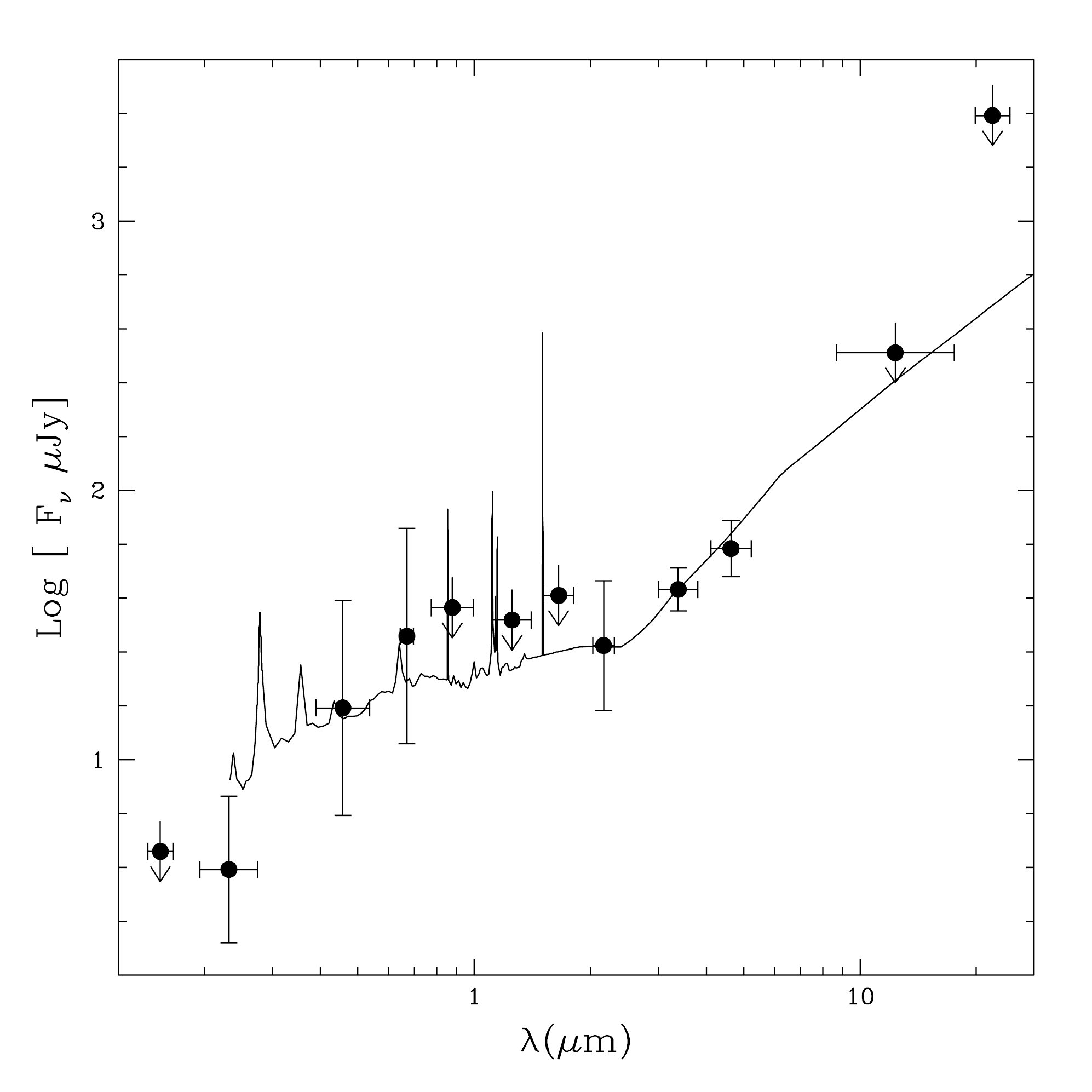}
\caption{\label{fig:x19sed}
{Spectral Energy Distribution of X19. From left to right, in order of increasing wavelength, the data points are: GALEX far-UV and near-UV; POSS-II $B_j$, $R_c$, and $I_c$; $J$, $H$, and $K_s$ from B07 and O10; WISE channels 1 -- 4. The best-fitting QSO model, a QSO1 from \citet{swire} at $z = 1.29$, is shown.
}
}
\end{figure}

We performed spectral energy distribution (SED) fitting using the UV and OIR data for X19. A \chisq\ minimization using a library of 726 galaxy and QSO models, and 254 stellar models (including 100 brown dwarf models), was performed using the Le PHARE software \citep{lephare}. The best-fitting galaxy or QSO template, a QSO1 from the SWIRE template library \citep{swire} redshifted to $z = 1.29^{+0.16}_{-0.07}$, had $\rchisq =1.47$; the best-fitting stellar model, a brown dwarf from \citet{chabrier:00}, was a poor match to the data ($\rchisq = 24$). In Fig.~\ref{fig:x19sed}, we show the SED of X19 along with the multiwavelength UV/OIR photometry.

The X-ray light-curve and spectrum, and the positional coincidence of the X-ray and UV/OIR sources, make for a convincing argument that X19 is a QSO at $z = 1.29$.
The offset between X19 and 5T7 must then be due to one of two possibilities: either 5T7 is not at $z = 1.29$, in which case X19 and its multiwavelength counterpart are unassociated with the radio transient, which remains unidentified in X-rays; or 5T7 is at $z = 1.29$, in which case it appears to be offset by $47 \pm 27$\,kpc from the X-ray position. We discuss this source further in Section~\ref{sec:x19prog}.

\section{Transient Progenitors}

It is important to note here that it is unlikely that all the radio sources belong to the same population.
Three, or maybe four of the steady sources appear to be associated with galaxies in the OIR images. These are likely AGNs or starburst galaxies. The remainder may be Galactic or extragalactic, but are likely distant or highly obscured. 

The transient sources with no counterparts may be very distant, highly obscured, or intrinsically faint in quiescence. Again, this is true whether they are Galactic or extragalactic, although in the latter case, as argued by O10, we would still expect to see more counterparts.

The two transient sources with clear OIR counterparts, and perhaps the two with possible counterparts, may, as noted by B07, be supernovae, although if so, they are unusually radio-luminous. For these four sources, no X-rays were seen in our \chandra\ image. This would not be particularly surprising if these sources are supernovae, particularly since the X-ray observations were taken between 11 and 27 years after the radio transients were seen. It has also been recently suggested \citep{nakar:11} that the radio emission in the two cases with clear counterparts may be from neutron star -- neutron star or neutron star -- black hole mergers. 

Of the six transient sources with no OIR identification, two have possible X-ray counterparts in our \chandra\ image (although one is a marginal detection and the other is offset too much for the association to be certain), and four have only upper limits. The two sources with ``possible'' OIR IDs as noted by B07 (which we argue are also unidentified) also have only upper limits in our \chandra\ data.

The count rates of the upper limits are similar to those for the detections, however, so it is conceivable that all the transients have similar count rates, of order $10^{-4}$~counts~s$^{-1}$. They may or may not be due to a single class of progenitor. It is worth noting that the only short-duration transient with a possible X-ray counterpart (5T7/X19) is by far the brightest of the transients in the radio (Table~\ref{tab:xflux}), which could indicate that it is a member of a different class of objects to the others.

\begin{deluxetable}{llllll}
\tablewidth{0pt}
\tabletypesize{\scriptsize}
\tablecaption{\label{tab:xflux} X-ray Flux Densities for Radio Transients}
\tablehead {
\colhead{VLA ID} & 
\colhead{\chandra} &
\colhead{\fullband\ rate} &
\colhead{$S_{\fullband}$ (NS)\tablenotemark{a}} &
\colhead{$S_{\fullband}$ (FS)\tablenotemark{b}} &
\colhead{$S_{5\,{\rm GHz}}$\tablenotemark{c}} \\
 & 
\colhead{ID} &
\colhead{(counts s$^{-1}$)} &
\colhead{($10^{-15}$~\esc)} &
\colhead{($10^{-15}$~\esc)} &
\colhead{(mJy)} 
}
\startdata
5T1 (RT\,19840502) & \nodata & $< 2.00 \times 10^{-4}$ & $< 6.17$ & $< 1.92$ & 0.45\\
5T2 (RT\,19840613) & \nodata & $< 2.51 \times 10^{-4}$ & $< 7.74$ & $< 2.41$ & 0.57\\
5T3 (RT\,19860115) & \nodata & $< 1.09 \times 10^{-4}$ & $< 3.36$ & $< 1.05$ & 0.37\\
5T4 (RT\,19860122) & \nodata & $< 2.00 \times 10^{-4}$ & $< 6.17$ & $< 1.92$ & 1.59\\
5T5 (RT\,19920826) & \nodata & $< 1.17 \times 10^{-4}$ &  $< 3.61$& $< 1.12$ & 0.64\\
5T6 (RT\,19970528) & \nodata & $< 2.67 \times 10^{-4}$ & $< 8.23$ & $< 2.57$ & 1.73\\
5T7 (RT\,19990504) & X19          & $ 8.20 \times 10^{-4}$ & $   25.3\tablenotemark{d}$& $7.88\tablenotemark{d}$ & 7.04 \\
8T1 (RT\,19970205) & \nodata & $< 1.76 \times 10^{-4}$ & $< 5.43$ & $< 1.69$ &2.23 \\
5L1 (RT\,19870422) & \nodata & $< 4.29 \times 10^{-4}$ & $< 13.2$ & $< 4.12$& 0.51\\
8L1 (RT\,20010331) & X124   & $ 1.78 \times 10^{-4}$ & $  5.49$ & $ 1.71$& 0.70\\
\enddata
\tablenotetext{a}{Unobscured flux for a basic neutron star model (blackbody with $kT = 0.1$\,keV)}
\tablenotetext{b}{Unobscured flux for a basic flare star / brown dwarf model (thermal bremsstrahlung with $kT = 1.0$\,keV)}
\tablenotetext{c}{From column~4 of Tables~2 and 3 of B07. For the 8\,GHz transients, we assume a flat spectral index between 5 and 8\,GHz, consistent with the discussion in O10.}
\tablenotetext{d}{The X-ray source appears to be an AGN, and values from more sophisticated spectral fits for this source from Table~\ref{tab:x19} (see also Section~\ref{sec:specfit}) should be used. The X-ray source may not be physically associated with 5T7 (Section~\ref{sec:x19sed}).}
\end{deluxetable}

We can achieve an improvement in sensitivity by ``stacking'' the X-ray images at the radio positions of the six sources with no OIR identification and only X-ray upper limits (5T1, 5T3, 5T4, 5T5, 5T6, and 8T1; see Table~\ref{tab:radio}). Stacking is a well-established \citep[\eg,][]{zibetti:05,devries:07} method for combining data from individual sources in order to study the statistical properties of objects which are undetected in single images. Prior information about positions at a frequency where sources are individually detected (in this case, the radio positions) is used to create ``cut-out'' images centered on those positions at a second frequency (here, from our X-ray data) where sources may not be individually detected. If $N$ images each with rms $\sigma$ are combined, then the noise level of the stacked image will decrease approximately as $\sigma / \sqrt{N}$. This allows sources below the original detection threshold to be studied. Stacks can also be created at ``blank sky'' positions offset from those of the known sources, in order to determine the background noise level in the absence of any additional emission present at the source positions.

We created a stack from the mean of six X-ray images aligned at the radio positions, and eight blank sky stacks; four at positions offset 30\arcsec\ in the cardinal directions, and four offset 42\arcsec\ in the ordinal directions from the radio positions. In other words, we created stacked images at positions on a $3 \times 3$ grid, with spacing 30\arcsec, centered on the radio positions.

We measured \chandra\ counts in a 5\arcsec-radius aperture at the center of each stacked image. The stacks offset from the radio positions will be dominated by background photons, whereas the stack centered at the radio positions will contain any additional flux from the radio sources, if present (assuming proper motions smaller than a few arcseconds over the time between the VLA and \chandra\ observations). The mean and rms of the eight background values was $7.09 \pm 0.89$ counts. There were 7.93 counts in the aperture in the stack made at the radio positions. So there is less than $1\sigma$ excess of X-ray counts at the radio positions of the non X-ray detected sources.

\subsection{Models, Fluxes, and Spectra}

We used the WebPIMMS v4.2 software\footnote{\url{http://cxc.harvard.edu/toolkit/pimms.jsp}} to convert the count rates into flux densities (for the two \chandra\ sources near to B07 transients) or upper limits (for the eight transients with no X-ray counterpart). The calculations are somewhat dependent on the model used for the X-ray emission. 

\subsubsection{Simple Models}

We chose to model two cases: a neutron star, and a flare star. The neutron star case was modelled as a blackbody with energy $kT = 0.1$\,keV, since isolated neutron stars older than $\sim 10^5$\,y have temperatures $\lesssim 0.1$\,keV \citep{yakovlev}. We caution, however, that the only isolated neutron stars which have been observed to date are relatively young, and may not be representative of older populations. Slow accretion onto magnetic neutron stars with fields of $\sim 10^{12}$\,G can cause $\gtrsim 10$\%\ of the X-ray luminosity to be emitted from broadened cyclotron emission features in the hard X-ray ($\sim 10$\,keV) regime \citep{nelson:93}. Other authors, however \citep[\eg][]{zane:00} note that although the spectrum for such objects is slightly harder than thermal, it is not markedly different from a blackbody. 

\label{sec:flaremodel}For the flare star we used a bremsstrahlung model with $kT = 1.0$\,keV. Both flare and post-flare spectra are often modelled in the literature using two-temperature models \citep[\eg][]{osten:10}. Since we have no reason to believe that the progenitors were in outburst when observed with \chandra\ (in contrast to when they were seen to flare in our VLA observations), the post-flare model is appropriate here. We choose a 1\,keV single temperature model, which is a valid approximation for both flare stars \citep{bildsten:00} and brown dwarfs \citep{imanishi:01,preibisch:05}. Blackbody, Raymond-Smith, and thermal bremsstrahlung models give similar fluxes (to within 30\%) in the \fullband\ energy range. We choose a thermal bremsstrahlung model.

More detailed spectral information could be used to constrain two-temperature fits or non-thermal components for both the neutron star and flare star cases, but since our sources are mostly undetected or barely detected, a simpler approach is warranted.

In both the neutron star and flare star cases, we assumed a Galactic \hi\ column density of $3 \times 10^{20}$\,cm$^{-2}$ \citep{hi}. We show the resulting flux densities and limits in Table~\ref{tab:xflux}.

For source X124/8L1, which was detected with only 8.8 counts, we have insufficient data to perform spectral modeling. However, we can use the $HR$ constraints ($0.38 \pm 0.44$) to constrain progenitor models. We used WebPIMMS to compute hardness ratios for blackbodies with a range of energies, and we find $kT > 0.8$\,keV at 90\%\ confidence. The best fit $kT = 1.4$\,keV.\label{sec:hr}

\subsection{Luminosities}

The X-ray luminosities of the transients (ignoring $k$-corrections if they are extragalactic) are 
\begin{equation}\label{eqn:xlum}
L_{\fullband} = 1.2 \times 10^{29} \left(\frac{S_{\fullband}}{{10^{-15}~}{\rm erg~s^{-1}~cm^{-2}}}\right) \left(\frac{d}{\rm 1\,kpc}\right)^2~{\rm erg~s^{-1},}
\end{equation}
where $S_{\fullband}$ is the model-dependent X-ray flux or upper limit from Table~\ref{tab:xflux} or Table~\ref{tab:x19}, and $d$ is the luminosity distance to the progenitor in kpc.

The radio luminosities of the transients are 
\begin{equation}\label{eqn:rlum}
L_{5\,{\rm GHz}} = 1.2 \times 10^{18} \left(\frac{S_{5\,{\rm GHz}}}{\rm 1\,mJy}\right) \left(\frac{d}{\rm 1\,kpc}\right)^2~\esh,
\end{equation}
where $S_{5\,{\rm GHz}}$ is the transient radio flux density at 5\,GHz (Table~\ref{tab:xflux}), or the upper limit (from B07) in quiescence (typically a few tens of \ujy), and $d$ is the luminosity distance to the progenitor in kpc. 

\subsection{Isolated Old Galactic Neutron Stars}

O10 suggest that the B07 transients may be due to isolated old Galactic neutron stars at distances between $\sim 1$ -- 5\,kpc. This implies X-ray luminosities around $10^{29}$ -- $10^{31}$\,erg~s$^{-1}$ for X124, and similar values as upper limits for the X-ray non-detections. This is consistent with the typical X-ray luminosities, $\lesssim 10^{31}$\,erg~s$^{-1}$, of old neutron stars accreting from the interstellar medium \citep[ISM;][]{treves:00,ostriker:70}. This luminosity depends on the accretion rate from the ISM, which in turn depends on the ISM density and the neutron star space velocity, and also on the field strength and spin rate of the neutron star. It also depends on the details of Bondi accretion in the presence of magnetic fields, which are not well understood \citep{toropina:03,toropina:05,arons:76,arons:80}.

However, although the measured X-ray fluxes and upper limits are consistent with a population of isolated neutron stars at distances between $\sim 1$ -- 5\,kpc, the hardness ratio measurements for X124 (Section~\ref{sec:hr}) exclude a simple one-component model with $kT = 0.1$\,keV, the expected energy for neutron stars. Since X19 is clearly not a neutron star, and X124 is the only other source for which we can compute hardness ratios, we cannot extrapolate this conclusion to the remainder of the X-ray sources. Additionally, as previously noted, accreting neutron stars may also have non-thermal components, and the spectra of old isolated neutron stars may not be identical to that of younger cooling neutron stars. Also, as previously noted, X124 may in fact be a spurious detection.

\citet{rutledge:03} note that optical luminosities of isolated neutron stars are around 5 orders of magnitude fainter in the optical than in the X-ray. Our X-ray fluxes of $\lesssim 10^{-14}$\,\esc\ would correspond to $\lesssim10^{-19}$\,\esc\ optical fluxes for this model, or $R \sim 35$ --- out of reach even for the planned James Webb Space Telescope. Accretion can cause brightening by a factor of a few in the optical \citep{zane:00}, but not likely enough to make our objects detectable.

\subsection{Magnetars}

Another class of potential neutron star progenitors discussed by B07 and O10 are soft gamma-ray repeaters (SGRs) or magnetars. These are attractive as potential progenitors because they are known to exhibit dramatic flares \citep{gaensler:05}, and would have no OIR counterparts in quiescence in our data. The X-ray emission from magnetars almost always has a blackbody component with $kT \sim 0.5$\,keV, and sometimes also has a power-law component \citep{axp}. Our X-ray data for X124 are not sensitive enough to constrain multi-component fits.

Typical magnetar X-ray luminosities are $\sim 10^{33}$\,erg~s$^{-1}$, so our sources would have to be at $\gtrsim 30$\,kpc, or underluminous if closer, to explain their non-detection or marginal detection in X-rays. As noted by B07 and O10, the surface density of magnetars is too low to explain all the B07 transients as being due to this class of object.

\subsection{Flare Stars}

Low mass stars and brown dwarfs are numerous, optically faint, and active 
at radio wavelengths in 5 to 30\% of cases \citep{berger:06}.  
M-dwarfs at distances as large as 1\,kpc
have been identified as the dominant contribution to optical
transient rates \citep{becker:04,kulkarni:06}. 

In an extreme case, the dMe star
EV Lac was observed to undergo 
a factor of $\sim 150$ flare to a radio luminosity
$\sim 10^{15}$\,\esh, 
with a rise time of minutes and a 
decay time of hours \citep{osten:05}.
Flare star (\ie, dMe star) radio luminosities in quiescence are generally no brighter than $\sim 10^{15}$\,\esh\ \citep{gudel:02}, implying even extreme flares ought to be no brighter than $\sim 10^{17}$\,\esh. The radio luminosities of our transients (Equation~\ref{eqn:rlum}; Table~\ref{tab:xray}) are $L_{\rm 5\,GHz} \sim 10^{18} (d/{\rm 1\ kpc})^2$\,\esh. So if the progenitors are flare stars, they must be closer than 1\,kpc, or else the transients must be due to unusually dramatic events. 

From the X-ray -- radio relation \citep{gudel:02}, our quiescent X-ray flux limits imply quiescent radio flux
densities of $\lesssim 1$\,\ujy\ (consistent with the non-detection of the transients in 
the deep radio image of B07 to depths of a few tens of \ujy). The inferred X-ray luminosities (Equation~\ref{eqn:xlum}) are $\lesssim 10^{30}~{\rm erg~s^{-1}}$ for progenitors at 1\,kpc, plausible for flare stars \citep{gudel:02}. Predicting X-ray luminosities during outburst from the peak radio luminosities of the transients (assuming the X-ray -- radio relation holds during outbursts), we obtain X-ray luminosities of $L_X \sim 10^{32} (d/{\rm 1\ kpc})^2 {\rm\ erg\ s^{-1}}$.
 
However, the non-detection of the transients to $K = 20.4$\,mag by O10 implies that if they are due to M5 stars, they must be at $d \gtrsim 1200$\,pc \citep{patten:06}. So unless the radio flares are unusually luminous, this implies that if flare stars are the progenitors of the transients, they must be of later spectral type than M5. 
 
O10 suggest that brown dwarfs at typical distances of around 200\,pc are plausible progenitors for the B07 transients, although they note that we would expect the radio emission to be strongly circularly polarized, in contrast to the low levels of polarization observed by B07. If they have spectral types as late as T8, they would be too faint to be detected by O10 if at distances $\gtrsim 50$\,pc \citep{patten:06}. 
Our X-ray fluxes imply luminosities of $\lesssim 10^{27}$\,erg~s$^{-1}$ at 30\,pc. This is a typical luminosity for quiescent emission from brown dwarfs \citep{preibisch:05}; sources with non-detections in our X-ray data could be intrinsically less luminous, more obscured, or further away.

\subsection{Proper Motion of 8L1 (RT\,20010331) / X124}

As noted in Section~\ref{sec:proper2}, the offset between X19 and 5T7 cannot be explained by proper motion. There is no UV/OIR counterpart to X124, however, so it is plausible that a proper motion of $\mu = 0.18 \pm 0.12$~arcsec~yr$^{-1}$ is responsible for the $1\farcs70 \pm 1\farcs09$ offset between X124 and 8L1.

We can use the 90\%\ confidence limits for the proper motion, along with plausible ranges of tangential velocity, $v_t = 4.74 \mu d$\,km~s$^{-1}$, for different classes of progenitors, to infer plausible ranges of distance, $d$ (in parsecs) to X124/8L1. A typical low mass star with $v_t \lesssim 20$\,\kms\ would have to be closer than $\sim 60$\,pc. Some dwarf stars have tangential velocities $> 100$\,\kms, however \citep{faherty:09}, so X124/8L1 could plausibly be a high proper motion object at $\lesssim 300$\,pc.  This would appear to marginally rule out an M star progenitor for this source (since, as discussed above, M stars should have been seen by O10 unless at $\gtrsim 1$\,kpc), but the proper motion is consistent with a brown dwarf progenitor.

The measured proper motion is also not unreasonable for a neutron star at $\sim 100$\,pc \citep{ofek:09}, particularly if the source is in the high-velocity tail of the distribution. A neutron star with tangential velocity $v_t \lesssim 3000$\,\kms\ \citep{cordes:98,fg:06} could explain the proper motion of X124/8L1 if at $\lesssim 9$\,kpc. The lower limit for distances of old isolated neutron star progenitors from the OIR non-detections ($\gtrsim 1$\,kpc), as noted above, implies that the tangential velocity must be $\gtrsim 330$\,\kms.

For both the flare star and neutron star cases, relatively small systematic position offsets could reduce the size of the measured proper motion, bringing upper limits on the distance from transverse velocities into better agreement with lower limits from to the $K$-band non-detections by O10. If the faint X-ray detection is not in fact due to an astronomical source, we cannot constrain the proper motion of the progenitor.

\subsection{Progenitor of 5T7 (RT\,19990504)\label{sec:x19prog}}

The X-ray luminosity of X19 is $2 \times 10^{44}$\,\esec, typical for X-ray selected AGNs \citep{lusso:10}.

The non-detection of X19 to $S_{5\,GHz} = 0.117$\,mJy in the deep radio image of B07 implies that the quiescent radio luminosity of the quasar is $L_{5\,GHz} \lesssim 10^{31}$\,\esh. Radio loud AGNs are defined as those having radio to optical luminosity ratio $R \geq 10$, where $R =  L_{5\,GHz} / L_B$ \citep{kellerman:89}. Our estimate of the B-band flux density, $S_{B_j} \sim 0.016$\,mJy, implies  $R \lesssim 7$ from the deep radio data, so the AGN is not radio loud in quiescence.

As noted above, the facts that X19 appears somewhat extended due to off-axis broadening of the \chandra\ PSF, and that the VLA D-array radio beam is rather large for 5T7, suggest that the uncertainty in the position may be somewhat higher than the formal uncertainties. If, in fact, 5T7 is associated with the AGN, this would imply that the AGN brightened by a factor of $\gtrsim 10$ (see B07, Figure~4) to $L_{5\,GHz} \sim 7 \times 10^{32}$\,\esh, and then faded back by a factor of $\gtrsim 10$ into undetectability on a timescale of days (and been brighter when detected as a transient by a factor $\gtrsim 60$ relative to its mean luminosity on a timescale of 22 years; see B07, Table 2). During the flare, $R \sim 440$, so the AGN would have changed from radio-quiet to radio-loud in this scenario.

It is more likely, however, due to the positional offset, that the radio source is unassociated with the nuclear X-ray source. If the radio transient is also at $z = 1.29$, the positional offset implies it is $47 \pm 27$\,kpc from the nucleus. In this case, the $L_{5\,GHz} \sim 7 \times 10^{32}$\,\esh\ radio transient could plausibly be due to a GRB afterglow in the outer regions or halo of the AGN host galaxy, although it would be among the brightest yet seen \citep{frail:web}. 

If in fact 5T7 is a chance association with X19, then the radio transient remains undetected in X-rays and should be considered along with the other non-detections as a possible brown dwarf or neutron star.

\subsection{Galaxy Cluster\label{sec:cluster}}

For source 5S1, apparently associated with a galaxy cluster, we used WebPIMMS to model the X-ray source X29 as a power law with $\Gamma =  1$ (optically thin bremsstrahlung) at $z = 0.25$, and a Galactic \hi\ column density of $3 \times 10^{20}$\,cm$^{-2}$, which results in a luminosity (extrapolated to rest frame 0.05 -- 40\,keV) of $\sim 5 \times 10^{43}$\,erg\,s$^{-1}$. Following \citet{reiprich:02}, we infer a mass of $\sim 3 \times 10^{14}$\,\msun\ for the cluster. The cluster could be part of a structure associated with Abell\,2047 (see B07), which is centered 10\farcm6 away, or $\sim 2$\,Mpc if both are at $z \sim 0.25$, perhaps explaining the fact that several of the B07 sources seem to be at this redshift.

\section{Conclusions}

Of the ten B07 transients, eight are undetected in X-rays; one is associated with an X-ray source of marginal ($2.4\sigma$) significance, which may be due to cosmic rays or particle background; and one may be associated with a QSO or its host galaxy at $z = 1.29$, or may be a spurious positional coincidence. Because of the uncertainty regarding these two associations, it is plausible that all of the B07 transients remain undetected in our X-ray images.

The measured X-ray flux for the transient with a marginal X-ray detection (X124/8L1), and the upper limits for the undetected transients, are consistent with the X-ray luminosity expected for brown dwarfs at distances $\sim 100$\,pc, one of the possible progenitors discussed by O10. The small offset between the X124 and 8L1 positions could plausibly be due to the proper motion of such an object.

If instead the transients are due to flaring M stars, most of the radio transients ought to have been detected in quiescence in both the X-ray and OIR images, unless the flares were extreme events (not out of the question, since we know the radio flux densities increased by a factor $\gtrsim 100$) from progenitors at distances $\gtrsim 1$\,kpc. 

The X-ray fluxes are also consistent with the suggestion of O10 that the transients could be isolated neutron stars. The radio and OIR fluxes are consistent with this picture as well. If X124 is truly a counterpart to 8L1, hardness ratio constraints appear to rule out a simple one-component spectral model, but there are insufficient data to attempt a more sophisticated fit with additional thermal or non-thermal components.

It is possible that the B07 transients are a mixture of some of the above classes of object, or some of them may be due to some unknown class of transient. 

If the progenitors have X-ray fluxes slightly fainter than our sensitivity limit, deeper observations would be expected to detect more of the transients in X-rays, as well as improving the significance of the fainter detection (or rejecting it as a real source). X-ray or radio monitoring campaigns lasting for several months would have a good chance of catching these sources in outburst since for the most likely classes of progenitor, flares are likely to repeat. Near- or mid-infrared data would also be useful to confirm or refute the brown dwarf hypothesis.

\acknowledgments
Support for this work was provided by the National Aeronautics and Space Administration through \chandra\ Award Number GO0-11152X issued by the \chandra\ X-ray Observatory Center, which is operated by the Smithsonian Astrophysical Observatory for and on behalf of the National Aeronautics Space Administration under contract NAS8-03060.

This research made use of Montage, funded by the National Aeronautics and Space Administration's Earth Science Technology Office, Computation Technologies Project, under Cooperative Agreement Number NCC5-626 between NASA and the California Institute of Technology. Montage is maintained by the NASA/IPAC Infrared Science Archive.

We thank Gregg Hallinan for helpful discussions regarding brown dwarfs. We thank the anonymous referee for a careful reading of the paper, and for constructive comments.

\end{document}